\documentclass[
reprint,
superscriptaddress,
 amsmath,amssymb,
 aps,
prb,
]{revtex4-2}

\usepackage{graphicx}
\usepackage{dcolumn}
\usepackage{bm}
\usepackage[hidelinks]{hyperref}

\usepackage{color}

\newcommand{\im}{\mathrm{i}}

\begin{document}

\preprint{APS/123-QED}

\title{Charge-density-wave melting in the one-dimensional Holstein model}

\author{Jan Stolpp}
 \affiliation{Institut for Theoretical Physics, Georg-August-Universit\"at G\"ottingen, D-37077 G\"ottingen, Germany}

\author{Jacek Herbrych}
 \affiliation{ Department of Theoretical Physics, Faculty of Fundamental Problems of Technology, Wroc\l aw University of Science and Technology, 50-370 Wroc\l aw, Poland}
\author{Florian Dorfner}
 \affiliation{Department of Physics, Arnold Sommerfeld Center for Theoretical Physics, Ludwig-Maximilians-Universit\"at M\"unchen, D-80333 M\"unchen, Germany}

\author{Elbio Dagotto}
 \affiliation{Department of Physics and Astronomy, The University of Tennessee, Knoxville, Tennessee 37996, USA}
 \affiliation{Materials Science and Technology Division, Oak Ridge National Laboratory, Oak Ridge, Tennessee 37831, USA}

\author{Fabian Heidrich-Meisner}
 \email{heidrich-meisner@uni-goettingen.de}
 \affiliation{Institut for Theoretical Physics, Georg-August-Universit\"at G\"ottingen, D-37077 G\"ottingen, Germany}

\date{\today}

\begin{abstract}
We study the Holstein model of spinless fermions, which at half filling exhibits a quantum phase transition from a metallic Tomonaga-Luttinger liquid phase to an insulating charge-density-wave (CDW) phase  at a critical electron-phonon coupling strength. In our work, we focus on the real-time evolution starting from two different types of initial states that are CDW ordered: (i) ideal CDW states with and without additional phonons in the system and (ii) correlated ground states in the CDW phase. We identify the mechanism for CDW melting in the ensuing real-time dynamics and show that it strongly depends on the type of initial state. We focus on the far-from-equilibrium regime and emphasize the role of electron-phonon coupling rather than dominant electronic correlations, thus complementing a previous study of photo induced CDW melting [H. Hashimoto and S. Ishihara, Phys. Rev. B {\bf 96}, 035154 (2017)]. The numerical simulations are performed by means of matrix-product-state based methods with a local basis optimization (LBO). Within these techniques, one rotates the local (bosonic) Hilbert spaces adaptively into an optimized basis that can then be truncated while still maintaining a high precision. In this work, we extend the time-evolving block decimation (TEBD) algorithm with LBO, previously applied to single-polaron dynamics,  to a half-filled system. We demonstrate that in some parameter regimes, a conventional TEBD method without LBO would fail. Furthermore, we introduce and use a ground-state density-matrix renormalization group method for electron-phonon systems using local basis optimization. In our examples, we account for up to $M_{\rm ph} = 40$ bare phonons per site by working with $O(10)$ optimal phonon modes.
\end{abstract}

\maketitle


\section{\label{sec:intro}Introduction}

Pump-probe experiments have become a popular setup to study ultrafast dynamics in solids (see, e.g., \cite{Schmitt-Science-2008,Tomeljak-PRL-2009,Ehrke-PRL-2011,Okamoto-PRB-2011,Stojchevska-Science-2014,Hu-NatMat-2014,Dal-Conte-NatPhys-2015,Giannetti-AiP-2016,Vogelsang-NatPhys-2018,Ligges-PRL-2018,Storeck-arxiv-2019}). In these experiments, photoinduced phase transitions between metallic and insulating states \cite{Okamoto-PRB-2011}, melting of charge-density-wave (CDW) or antiferromagnetic order \cite{Schmitt-Science-2008,Tomeljak-PRL-2009,Ehrke-PRL-2011}, or accessing metastable states \cite{Stojchevska-Science-2014} were investigated.
A prominent example is the observations of Ref.~\onlinecite{Hu-NatMat-2014} that were interpreted as photo induced enhanced superconductivity.
In the interpretation of experiments on ultrafast dynamics, the whole system is often treated as a collection of coupled subsystems \cite{Giannetti-AiP-2016}.
These include the electronic subsystem, lattice degrees of freedom (phonons), and possibly spin degrees of freedom. In the experiments, first electrons are optically excited into empty states and then their relaxation dynamics is monitored. Relaxation can occur via electronic interactions or via a coupling to bosons, i.e., either phonons or spin excitations.
Theoretical support is needed to understand the timescales and the bottlenecks for relaxation, and to determine which bosonic excitations are relevant.
In general, it is unclear whether the subsystems first relax and thermalize separately before reaching global equilibrium or whether all degrees of freedom are out-of-equilibrium throughout the transient dynamics. Moreover, the strength of phonon-mediated interactions could be affected in the transient dynamics \cite{Murakamu2017a,Kennes-Nature-2017}. Thus, a major task for theory is to understand such questions in simplified yet paradigmatic models. Many studies focused solely on electronic degrees of freedom (see, e.g., \cite{Moritz-CPC-2011,Werner-PRB-2012,Eckstein-PRL-2013,Lu-PRB-2015,Golez-PRB-2016,Kohler2018,Paeckel2019,Golez2019}), yet from the above it is clear that phonons need to be modeled as well \cite{Werner-EPL-2015,Kemper-AnPhys-2017,Murakami2017b}.

The Holstein model of spinless fermions in one dimension is a prototypical model to study electron-phonon coupled systems.
It hosts a variety of different phenomena driven by the electron-phonon coupling, especially polaron formation and a phase transition between a metallic and a CDW phase \cite{Barisic-EPJB-2008,Hirsch-PRB-1983}.
The rich phenomena present in the Holstein model and, in particular, its nonequilibrium dynamics are still actively discussed.
Studies of the latter in electron-phonon coupled systems are often restricted to single electrons (Holstein-polaron problem) \cite{Vidmar-PRB-2011,Fehske-PRB-2011,Golez-PRL-2012,Sayyad-PRB-2015,Dorfner-PRB-2015,Kloss-PRL-2019}.
However, having more than one electron in the system can lead to interesting collective phenomena already in equilibrium. One of the most prominent examples is the Peierls instability leading to an insulating CDW-ordered state in a half-filled electron band coupled to phonons.
Despite the challenges, efforts were made to study the real-time dynamics in the Holstein model at half filling \cite{Filipps-PRL-2012,Matsueda-JPSJ-2012,Hohenadler-PRB-2013,Werner-EPL-2015,Wall-PRA-2016,Hashimoto-PRB-2017}.

\begin{figure}[!tb]
\includegraphics[width=\columnwidth]{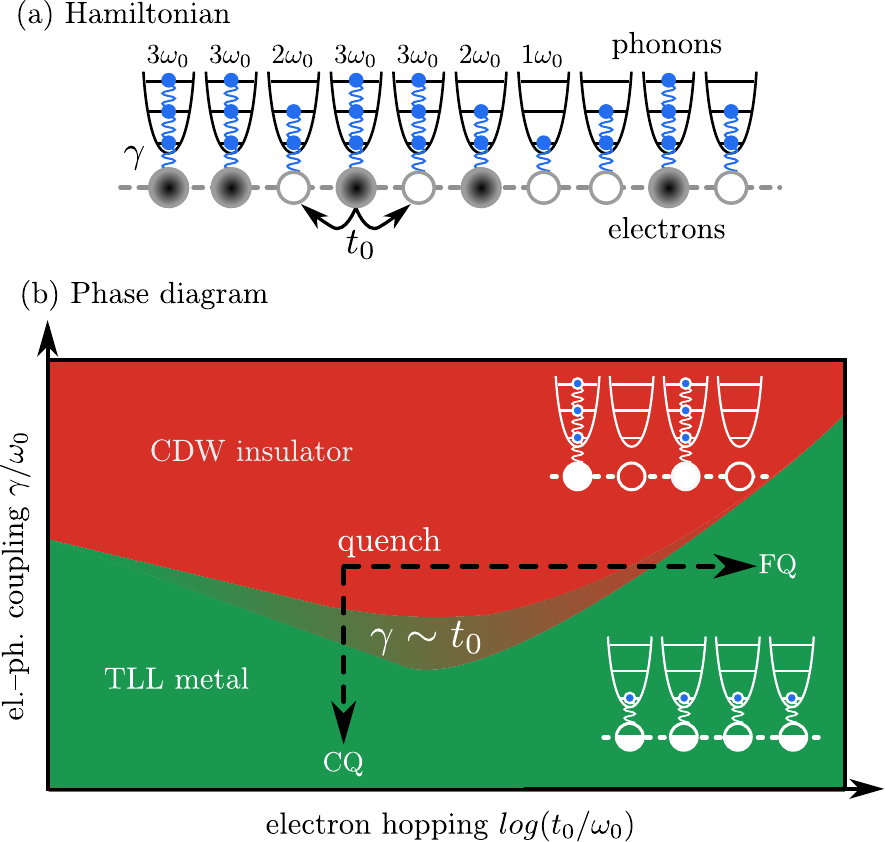}
\caption{\label{fig:holstein_sketch} (a) Sketch of the different terms in the Holstein model Eq.~\eqref{eq:hol-ham}. The fermions can hop from site to site with an amplitude $t_0$. If a fermion is on a particular site it can create or destroy phonon excitations at that site with a coupling strength $\gamma$. Every phononic excitation costs an energy $\omega_0$. (b) Sketch of the phase diagram of the half-filled Holstein model \cite{Bursill-PRL-1998,Creffield-EPJB-2005}. As the electron-phonon coupling $\gamma$ increases there is a phase transition from a metallic Tomonaga-Luttinger liquid phase (TLL) to an insulating charge-density-wave phase (CDW). The arrows represent the different quenches that we will investigate in Sec.~\ref{sec:quench}, i.e., frequency and coupling quench (FQ and CQ, respectively).}
\end{figure}

Perturbative approaches can give reliable results in the vicinity of the atomic limit, where the bandwidth of the electrons is much smaller than all other energy scales in the system \cite{Hirsch-PRB-1983} and also in the limit of small phonon energies \cite{Caron-PRB-1984}. For the intermediate regime, the so-called momentum average approach developed by Berciu and collaborators \cite{Berciu-PRL-2006,Goodvin-PRB-2006,Barisic-PRL-2007,Berciu-PRL-2007,Berciu-PRB-2007,Goodvin-PRL-2011} is argued to provide reliable analytical results for the Holstein-polaron problem in equilibrium.

A variety of different quantum Monte Carlo methods have been developed to investigate the Holstein model \cite{Hirsch-PRB-1983,McKenzie-PRB-1996,Kornilovitch-PRL-1998,Hohenadler-PRB-2004,Creffield-EPJB-2005,Goodvin-PRL-2011,Ohgoe-PRB-2014,Weber-PRB-2016}. For wave-function based methods, such as exact diagonalization (ED) or the density-matrix renormalization group (DMRG), electron-phonon systems are computationally very demanding.
These methods require that the local Hilbert spaces have a finite dimension, which is not the case for electron-phonon coupled systems.
The bosonic nature of the phonons and the fact that their number is not conserved makes the Hilbert space infinite-dimensional, irrespective of the system size.
Therefore, one has to introduce an {\it ad hoc} cutoff that limits the number of phonons per site.
This cutoff has to be chosen in such way that it does not affect the physics of the system and the quantitative reliability of the results.
Depending on the task at hand, this can render the problem unfeasible or at least very hard for wave-function based methods.

Several strategies were suggested to overcome the problem of large local Hilbert spaces \cite{Jeckelmann-RNC-2007}.
In the context of DMRG \cite{Schollwoeck-RMP-2005}, one can map the Holstein model to a lattice including pseudo sites for the phononic degrees of freedom where every pseudo site can host one phonon excitation \cite{Jeckelmann-PRB-1999,Jeckelmann-PRB-1998}.
As a result, the local Hilbert-space dimension is reduced; however, one introduces long-range hopping into the system.
Wei\ss e and Fehske \cite{Weisse-PRB-1998} used an inhomogeneous modified variational Lang-Firsov transformation to obtain an effective Hamiltonian including variational parameters that can be solved in a self-consistency loop including a Lanczos diagonalization \cite{Fehske-PRB-1995}.
In other approaches, one chooses basis states in such a way that the Hilbert space is not too big but still the essential physics is captured.
For instance, Bon\v ca {\it et al.} \cite{Bonca-PRB-1999} introduced diagonalization in a limited functional space.
In this approach, a set of dynamically important basis states is constructed by repeatedly applying parts of the Hamiltonian to an initial state.
This method takes advantage of the spatial correlations of electrons and phonons and is therefore especially well suited for studying single electrons on a periodic or infinite lattice \cite{Bonca-PRL-2000,Li-PRB-2010,Vidmar-PRB-2010,Vidmar-PRL-2011,Vidmar-PRB-2011,Golez-PRL-2012,Golez-PRB-2012,Dorfner-PRB-2015}.

In this work, we will use an approach called local basis optimization (LBO) introduced by Zhang {\it et al.} \cite{Zhang-PRL-1998}.
This approach is very flexible since it adaptively chooses the most important local basis states (called optimal modes) during the simulation by diagonalization of the single-site reduced density matrix. The ideas of Zhang {\it et al.} \cite{Zhang-PRL-1998} were first used in combination with exact-diagonalization techniques \cite{Zhang-PRL-1998,Zhang-PRB-1999,Nishiyama-EPJB-1999,Weisse-PRB-2000,Zhao-PRB-2005} and also with DMRG in its original formulation \cite{Bursill-PRB-1999,Friedman-PRB-2000,Friedman-JoPCM-2002,Bursill-PRB-2002,Barford-PRB-2002,Barford-PRB-2006,Wong-PRB-2008,Tozer-PRB-2014}.

Here, we will combine LBO with a time-dependent DMRG algorithm as well as with a ground-state DMRG algorithm in the matrix-product-state (MPS) formulation.
In these DMRG implementations, we choose the optimized basis in an unbiased way and fully adaptive to system size, system parameters, and boundary conditions.
The time-dependent version is based on the work  by Brockt {\it et al.}~\cite{Brockt-PRB-2015} to simulate the real-time evolution in the Holstein-polaron problem (see also Refs.~\cite{Schroeder-PRB-2016,Brockt-PRB-2017}).
In this work, we extend this algorithm to the Holstein model at half filling.
Our ground-state DMRG method combines the algorithm implemented by Guo {\it et al.}~\cite{Guo-PRL-2012} for spin-boson models (see also \cite{Bruognolo-PRB-2014,Blunden-Codd-PRB-2017,Bruognolo-PRB-2017}) with the subspace expansion introduced by Hubig {\it et al.}~\cite{Hubig-PRB-2015}.
The algorithm can be applied to arbitrary one-dimensional electron- or spin-phonon problems with local electron- or spin-phonon coupling.
Here, we use this algorithm to study the half-filled Holstein model.

In the first setup, we prepare the system in a product state where every second site is occupied by an electron and no phonons are present in the initial state.
We then perform a real-time evolution of this state for different parameter sets.
As we increase the coupling to the phonons we observe a transition from dynamics that is dominated by the electron hopping to dynamics that is strongly influenced by the coupling to the phonons. 
This includes a temporal self-trapping of the electrons for large electron-phonon coupling.
In the second setup, we prepare the system in a product state of small on-site polarons that form the CDW.
In this case, the real-time evolution can be understood by considering the renormalized hopping-matrix elements of the quasiparticles.
As a consequence, the dynamics at strong coupling is so slow that the initial state hardly changes over our accessible simulation times.
In the last setup, we prepare the system in the ground state of the CDW phase and then we perform quenches to the metallic phase. 
We observe that the short-time dynamics is dominated by the phonons when we decrease the coupling between electrons and phonons only.
However, if we decrease the phonon frequency compared to the electron bandwidth the short-time dynamics is dominated by the electron hopping, while the phonons respond very slowly to the quench.

The melting of charge-density-wave states in a one-dimensional electron-phonon coupled system was previously studied by Hashimoto and Ishihara \cite{Hashimoto-PRB-2017} using time-dependent DMRG simulations with a fixed cutoff in the local phonon number basis of $M_{\text{ph}} \leq 8$.
They study a Holstein model with an electronic interaction of the form $H_{\text{int}} = V\sum_l n_l n_{l+1} $ ($n_l=c^\dagger_l c^{\vphantom{\dagger}}_l$, $c^{\vphantom{\dagger}}_l$: fermionic annihilation operator at site $l$) and drive the system out of  equilibrium by applying a pulse.
Starting from the uncoupled limit of a vanishing electron-phonon coupling [$\gamma=0$, cf. Fig.~\ref{fig:holstein_sketch}(a)], they demonstrate that the CDW order parameter decays exponentially for $V>t_0$ [where $t_0$ is the electron hopping parameter; cf.~Fig.~\ref{fig:holstein_sketch}(a)].
Turning on electron-phonon interactions causes a slower decay due to the formation of polarons and thus a mass renormalization of the electrons.
The excess energies pumped into the system that were considered in \cite{Hashimoto-PRB-2017} are on the order of $\Delta E \lesssim  0.1 t_0 N$ above the ground-state energy, where $N$ is the number of fermions in the system.
In our work, we consider different initial states and we deliberately work in the regime of large quench energies $0.1 t_0 N \lesssim \Delta E\lesssim 8t_0 N$ to exemplify the capabilities of our local basis-approximation method.

The paper is organized as follows. In Sec.~\ref{sec:model}, we will revisit the Holstein model and its phase diagram at half filling. In Sec.~\ref{sec:num_methods}, we describe the different numerical methods used throughout this paper. In Sec.~\ref{sec:nonequ}, we present the results of our numerical simulations and in Sec.~\ref{sec:summary}, we give a summary.

\section{\label{sec:model}Holstein model of spinless fermions}

The Holstein model \cite{Holstein-I-AoP-1959,Holstein-II-AoP-1959} of spinless fermions describes a spin-polarized gas of electrons that locally couples to harmonic oscillators via the density of the electrons.
The harmonic oscillators model dispersionless phonons.
The Hamiltonian on a one-dimensional (1D) lattice can be written as
\begin{align}
H_{\rm Hol} = H_{\rm kin} + H_{\rm ph} + H_{\rm el-ph}\,, \label{eq:hol-ham}
\end{align}
where $H_{\rm kin}$ is the electron kinetic energy, i.e.,
\begin{align}
H_{\rm kin} = -t_0 \sum_{l=1}^{L-1} (c_l^\dagger c_{l+1}^{\vphantom{\dagger}} + h.c.)\,.
\end{align}
Here the $c_l^{\vphantom{\dagger}}$ [$c_l^\dagger$] are annihilation [creation] operators for spinless fermions and $t_0$ is the hopping parameter. 
$H_{\rm ph}$ is the purely phononic part defined as
\begin{align}
H_{\rm ph} = \omega_0 \sum_{l=1}^L b_l^\dagger b_l^{\vphantom{\dagger}}\,,
\end{align}
where $b_l^{\vphantom{\dagger}}$ [$b_l^\dagger$] are bosonic annihilation [creation] operators for phonons and $\omega_0$ is the phonon frequency. $H_{\rm el-ph}$ is the electron-phonon coupling part:
\begin{align}
H_{\rm el-ph} = -\gamma \sum_{l=1}^L n_l^{\vphantom{\dagger}} (b_l^\dagger + b_l^{\vphantom{\dagger}})\,, \label{eq:ham-el-ph}
\end{align}
where $n_l^{\vphantom{\dagger}} = c_l^\dagger c_{l}^{\vphantom{\dagger}}$ is the on-site density of the electrons and $\gamma$ is the electron-phonon coupling strength. The different parts of the Holstein Hamiltonian Eq.~\eqref{eq:hol-ham} are sketched in Fig.~\ref{fig:holstein_sketch}(a).
The total number of fermions $N = \sum_{l=1}^L  \langle n_l^{\vphantom{\dagger}}\rangle$ is conserved in the system while the number of phonons is not, as is evident from Eq.~\eqref{eq:ham-el-ph}. Throughout this paper, we express energies and times in units of the hopping parameter $t_0$ and $1/t_0$, respectively. Open boundary conditions are used within our numerical simulations.

In Fig.~\ref{fig:holstein_sketch}(b), we sketch the ground-state phase diagram of the half-filled Holstein model that was obtained by a combination of perturbative approaches, quantum Monte Carlo and DMRG methods \cite{Creffield-EPJB-2005,Bursill-PRL-1998,McKenzie-PRB-1996,Hirsch-PRB-1983}. For small values of the  coupling parameter $\gamma/\omega_0 \ll 1$, the system is in a (metallic) Tomonaga-Luttinger liquid (TLL) phase while for increasing coupling strength $\gamma/\omega_0$, there is a phase transition to a CDW phase for all values of the hopping parameter $t_0>0$.
The order parameter in the latter can be defined as the staggered density of the fermions in the system:
\begin{align}
\mathcal{O}_{\rm CDW} = \frac{1}{N} \sum_{l=1}^L (-1)^l \langle n_l \rangle \,.
\end{align}
In the metallic TLL phase, the density is homogenous $\langle n_l \rangle = 0.5 = \mathrm{const.}$ and therefore, the order parameter vanishes. On the other hand, $\mathcal{O}_{\rm CDW} \neq 0$ indicates the onset of the CDW phase, with a maximum value of $\mathcal{O}_{\rm CDW} = \pm 1$ in the limit $\gamma/t_0 \rightarrow \infty$.
This is strictly true in the thermodynamic limit, yet we will break the symmetry here by the choice of initial conditions or system size and boundary conditions.

A subtlety that emerges from using small odd system sizes is that the order parameter $\mathcal{O}_{\rm CDW}$ can be zero although the density is not completely uniform. This arises because there is one more odd site than there are even sites. However, this should not be concerning. Consider free spinless fermions on a lattice with odd system size $L$ and open boundary conditions. The number of fermions is $N = (L-1)/2$.
Then, in the ground state, the $N$ lowest single-particle eigenstates are occupied which also leads to a density profile that is not flat but has exactly $\mathcal{O}_{\rm CDW} = 0$. This effect becomes less 
pronounced as the system size is increased.

In the atomic limit $t_0 = 0$, the Holstein model can be diagonalized by performing a Lang-Firsov transformation \cite{Lang-SP-1963}.
In the ground state, fermions are localized at single sites and are accompanied by coherent states of phonons. 
All other sites that do not contain a fermion are free of any phonon.
For every fermion in the system, one gets a binding energy:
\begin{align}
\epsilon_b = \frac{\gamma^2}{\omega_0} \,,
\end{align}
and the ground-state energy is therefore $E_0 = - N \epsilon_b$. 
The ground state in this limit is highly degenerate since one can distribute the fermions arbitrarily in the system. It takes the form of a product state:
\begin{align}
| \psi_0 \rangle \propto  \left[  \prod_{l \in \{l_{\rm occ}\}} c_{l}^\dagger \ e^{\frac{\gamma}{\omega_0}b_{l}^\dagger} \right] | \emptyset \rangle _{\rm el}  | \emptyset \rangle_{\rm ph}\,,  \label{eq:gs-lang-firsov}
\end{align}
where $| \emptyset \rangle _{\rm el[ph]}$ is the vacuum state of the electrons [phonons] and $\{ l_{\rm occ}\}$ is the set of sites that are occupied.

Close to the atomic limit $t_0 \ll \gamma, \omega_0$ one can understand the phase transition from second-order perturbation theory \cite{Hirsch-PRB-1983}. One obtains an effective polaron hopping-matrix element,
\begin{align}
\tilde{t}_0 = t_0 e^{-\gamma^2/\omega_0^2} \label{eq:teff}
\end{align}
and an effective nearest-neighbor repulsion,
\begin{align}
\tilde{V} = 2 \frac{\tilde{t}_0^2}{\omega_0} \int_0^{\frac{\gamma^2}{2\omega_0^2}} dg\, \frac{e^{4g}-1}{g}\,.
\end{align}
The effective model can then be mapped to the spin-1/2 {\it XXZ} Hamiltonian and the phase transition at the isotropic Heisenberg point is reached at $\tilde{V}_c/2\tilde{t}_0 = 1$ \cite{Hirsch-PRB-1983}.

\section{Numerical methods}
\label{sec:num_methods}

\subsection{Ground-state DMRG with local basis optimization}
\label{sec:dmrg+lbo}

To calculate ground states of the half-filled Holstein model we use a single-site DMRG algorithm and combine this with a local basis optimization (LBO) \cite{Zhang-PRL-1998}.
In the first efforts to combine LBO with DMRG, the optimal modes were computed from small systems using exact diagonalization and then fed into larger systems (see, e.g., \cite{Bursill-PRB-1999,Friedman-PRB-2000,Bursill-PRB-2002})
or the optimal modes were computed from units larger than a single site (see, e.g., \cite{Wong-PRB-2008}).
The algorithm presented in \cite{Barford-PRB-2002} uses the original DMRG formulation \cite{White-PRL-1992} and is the closest to our implementation and the one of \cite{Guo-PRL-2012}, yet uses different environment-block DMRG basis dimensions depending on whether optimal-phonon mode optimization takes place or not.

The algorithm used in this work is an adaptation of the method described in \cite{Guo-PRL-2012} to electron-phonon systems combined with the subspace-expansion method (DMRG3S) \cite{Hubig-PRB-2015}.
Therefore, we use the abbreviation DMRG3S+LBO when referring to the method used in this work.

Consider a pure quantum state $\left| \psi \right\rangle$ of a lattice system that can be expanded in a product-state basis of $d$-dimensional local Hilbert spaces.
We start out by writing this state as a matrix-product state (MPS) in the standard fashion following Ref.~\onlinecite{Schollwoeck-AoP-2011}:
\begin{align}
\left| \psi \right\rangle = \sum_{\{ \sigma_l \}} a_{ \sigma_1 ... \sigma_L} \left| \sigma_1 ... \sigma_L \right\rangle = \sum_{\{ \sigma_l \}} M^{\sigma_1} ... M^{\sigma_L} \left| \sigma_1 ... \sigma_L \right\rangle\,,
\end{align}
where the $\sigma_l$ label the state in the local Hilbert space and $M^{\sigma_l}$ are matrices such that the matrix product yields  $M^{\sigma_1} ... M^{\sigma_L} = a_{ \sigma_1 ... \sigma_L}$(actually, the first matrix $M^{\sigma_1}$ and the last matrix $M^{\sigma_L}$ have to be a row and column vector, respectively, for the matrix product to yield a scalar). The sum runs over all possible values of $\sigma_1, ..., \sigma_L$. The full many-body Hilbert space has dimension $\mbox{dim}(\mathcal{H} ) = d^L$.

In principle, the dimension of the matrices $M^{\sigma_l}$ - the so-called bond dimension - also grows exponentially with the system size $L$ except at the edges of the system.
The success of MPS-based methods relies on the fact that ground states of short-range Hamiltonians in one dimension that have a gap to the excitation spectrum can be efficiently represented with matrices of a limited dimension that does not depend on the system size $L$ \cite{Laflorencie-PR-2016,Schollwoeck-AoP-2011,Eisert-RMP-2010,Hastings-JSM-2007}. This can be understood in the following way: divide the system into two parts and consider the reduced density matrix of one of these subsystems. If the spectrum of the reduced density matrix of the subsystems falls off fast enough, the state can be efficiently and accurately represented by considering just a limited part of the states in either one of the subsystems.
The area law of entanglement for the ground state of gapped short-range Hamiltonians in one dimension ensures a fast algebraic decay of the spectrum \cite{Hastings-JSM-2007}.
Therefore, it is enough to consider a finite dimension of the matrices $M^{\sigma_l}$ \cite{Schollwoeck-AoP-2011}.

Following Ref.~\onlinecite{Zhang-PRL-1998}, we now consider a special bipartition where we only look at one site. The local reduced density matrix at site $l$ is given by
\begin{align}
\rho_l = \underset{\underset{m \neq l}{\sigma_m}}{\rm tr}(\left| \psi \right> \left< \psi \right|)\,,
\end{align}
where the trace runs over all local degrees of freedom $\sigma_m$ that are not on site $l$. Diagonalizing this local density matrix we obtain
\begin{align}
\rho_l = U_l \Lambda_l U^\dagger_l\,,
\end{align}
where $\Lambda_l$ is a diagonal matrix with the eigenvalues of the local density matrix on the diagonal and $U_l$ is a local basis transformation from the original basis (in practice, this will most often be an occupation number basis) to the eigenbasis of the local reduced density matrix.

If the spectrum of the local reduced density matrix falls off fast enough it is advisable to rotate the original $M^{\sigma_l}$ of our MPS into the new $\tilde\sigma_l$ eigenbasis of the local reduced density matrix.
It is then sufficient to only keep that part of the eigenbasis with the largest eigenvalues of the local reduced density matrix without losing much of the information of the state \cite{Zhang-PRL-1998}.
Therefore, we introduce a truncated basis transformation $R^{ \tilde\sigma_l \sigma_l}$ that has dimensions $d_o \times d$ where $d_o < d$.
Here $R$ is identical to $U^\dagger$ with the exception that in $R$, we got rid of the $d-d_o$ rows of the matrix that correspond to the smallest eigenvalues of $\rho_l$. We then write the MPS as:
\begin{align}
| \tilde\psi \rangle = \sum_{\{ \sigma_l \}} (\tilde M^{\tilde\sigma_1}R^{\tilde\sigma_1 \sigma_1}) ... (\tilde M^{\tilde\sigma_L}R^{\tilde\sigma_L \sigma_L}) \left| \sigma_1 ... \sigma_L \right\rangle\,,
\end{align}
where:
\begin{align}
 \tilde M^{\tilde\sigma_l} = M^{\sigma_l} R^{\dagger \sigma_l \tilde\sigma_l}\,.
\end{align}

The rotation into an optimized local basis is motivated by the observation that the Holstein model Eq.~\eqref{eq:hol-ham} can be diagonalized in the atomic limit $t_0 = 0$ via a Lang-Firsov transformation as discussed in Sec.~\ref{sec:model}.
The ground state of the model will then take the form Eq.~\eqref{eq:gs-lang-firsov} where the sites that are occupied by an electron also contain a coherent state of phonons.
In order to represent this state accurately, large phonon occupations  need to be accounted for  such that the Hilbert space in the phonon occupation basis has to have a large dimension.
On the other hand, in the Lang-Firsov basis, a two-dimensional local Hilbert space is enough to represent the ground state: one state for a site occupied by an electron and one for an empty site.
That is, in the atomic limit keeping only one state per fermion occupation sector is sufficient to represent the ground state exactly.
Away from the atomic limit, keeping only $d_o \ll d$ states is still sufficient to accurately represent the ground state \cite{Zhang-PRL-1998,Zhang-PRB-1999} as we will see in the following.
In fact, Zhang {\it et al.} found numerical evidence that the spectrum of the local reduced density matrix falls off exponentially in ground states \cite{Dorfner-PRA-2016,Zhang-PRL-1998}, which seems to hold also in time-evolved states of the Holstein polaron model \cite{Dorfner-PRB-2015}.
Diagonalizing the local density matrix automatically finds the optimal basis to represent the state.
Manipulations on the MPS matrices $M^{\tilde\sigma_l}$ that we have to do during the DMRG sweeps become cheaper because of the reduced dimensionality when using the optimized local basis.
We stress that the outlined ansatz finds the optimized basis at every site adapted to the system parameters, boundary conditions, and also time during a time evolution.

\begin{figure}[!tb]
\includegraphics[width=\columnwidth]{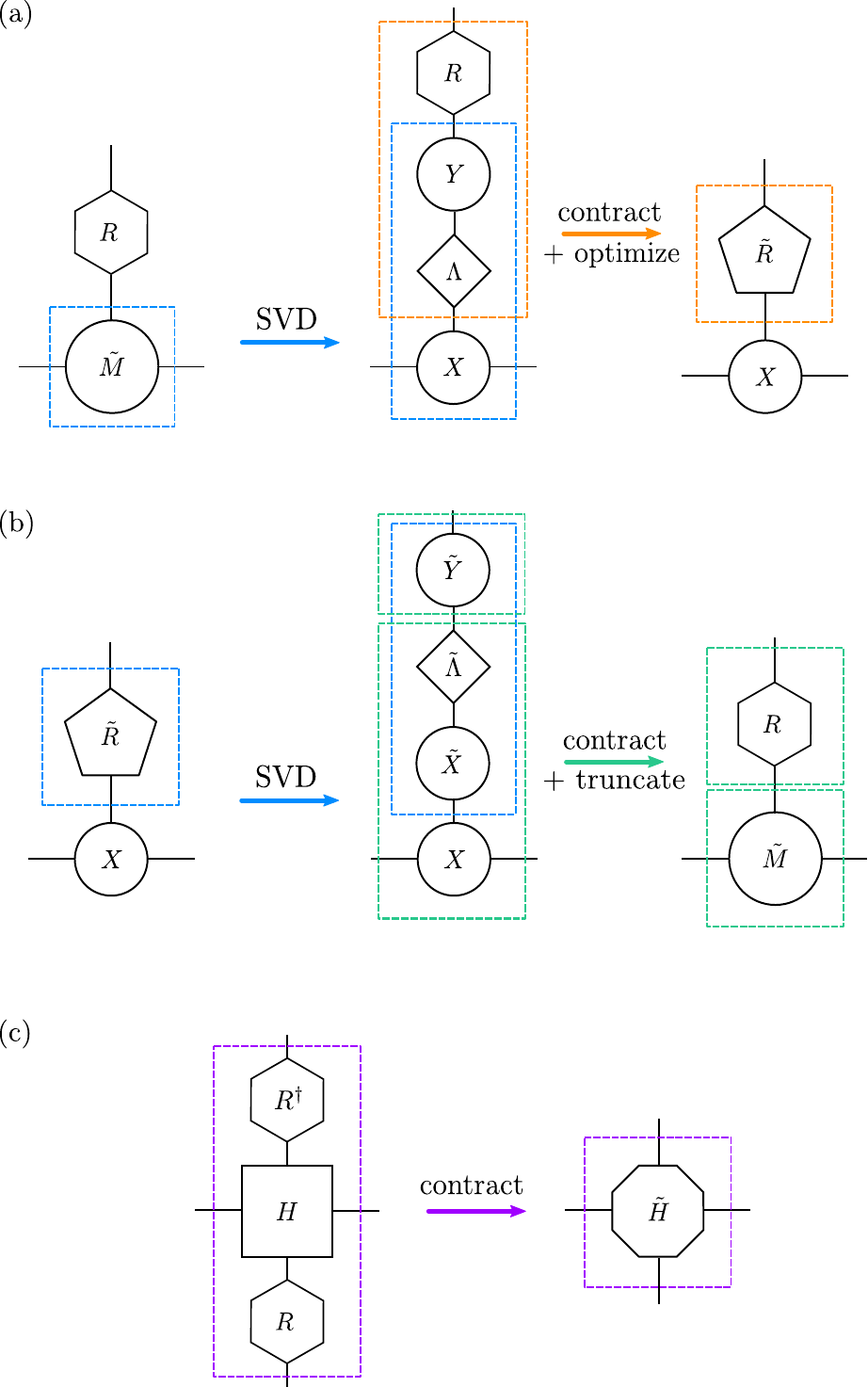}
\caption{\label{fig:sketch_dmrg_lbo} Different steps of the DMRG3S+LBO update. (a) Shift of the focus to the basis transformation tensor and optimization. (b) Shift of the focus back to the site tensor and truncation. (c) Transformation of the local part of the Hamiltonian MPO into the optimized basis (see also \cite{Guo-PRL-2012}).}
\end{figure}

We will now explain the basic steps of our algorithm. We consider an MPS in mixed canonical form where the MPS matrices are transformed into an optimal basis:
\begin{align}
\left| \psi \right\rangle = \sum_{\{ \sigma_l \}} \tilde A^{\tilde\sigma_1}_{a_0 a_1} R^{\tilde\sigma_1 \sigma_1} ... \tilde A^{\tilde\sigma_{i-1}}_{a_{i-2} a_{i-1}} R^{\tilde\sigma_{i-1} \sigma_{i-1}} \tilde M^{\tilde\sigma_{i}}_{a_{i-1} a_{i}} R^{\tilde\sigma_{i} \sigma_{i}} \nonumber \\
 \times  \tilde B^{\tilde\sigma_{i+1}}_{a_{i} a_{i+1}} R^{\tilde\sigma_{i+1} \sigma_{i+1}} ... \tilde B^{\tilde\sigma_{L}}_{a_{L-1} a_{L}} R^{\tilde\sigma_{L} \sigma_{L}} \left| \sigma_1 ... \sigma_L \right\rangle \,.\label{eq:mps-lbo}
\end{align}
Here and for the rest of the section a summation over all indices that appear twice is implied. The indices $a_0$ and $a_{L}$ are fixed dummy indices to standardize notation. The $\tilde A^{\tilde \sigma_l}_{a_{l-1} a_l}$ and $\tilde B^{\tilde \sigma_l}_{a_{l-1} a_l}$ are left- and right-normalized MPS tensors, respectively,
\begin{align}
\tilde A^{\dagger \tilde\sigma_l}_{a_l a_{l-1}} \tilde A^{\tilde\sigma_l}_{a_{l-1} a^\prime_l} &= \delta_{a_l a^\prime_l} \\
\tilde B^{\tilde\sigma_l}_{a^\prime_{l-1} a_l} \tilde B^{\dagger \tilde\sigma_l}_{a_l a_{l-1}} &= \delta_{a^\prime_{l-1} a_{l-1}}
\end{align}
such that the local reduced density matrix at site $i$ in the optimized basis can be written as
\begin{align}
(\rho_i)^{\tilde\sigma^\prime_i \tilde\sigma_i} =  \tilde M^{\tilde\sigma^\prime_i}_{a_{i-1} a_i} \tilde M^{\dagger \tilde\sigma_i}_{a_i a_{i-1}}.
\end{align}
The first step is to shift the focus of the state which is currently on the $\tilde M^{\tilde\sigma_i}$ tensor to the basis transformation tensor $R^{\tilde\sigma_i \sigma_i}$ such that the local reduced density matrix can be written in terms of only the $R^{\tilde\sigma_i \sigma_i}$ tensor instead of the $\tilde M^{\tilde\sigma_i}$ tensor. The different tensor manipulations that are necessary are depicted in Fig.~\ref{fig:sketch_dmrg_lbo}(a). We perform a singular value decomposition (SVD) of $\tilde M^{\tilde\sigma_i}$:
\begin{align}
\tilde M^{\tilde\sigma_i}_{a_{i-1} a_i} = X^\tau_{a_{i-1} a_i} \Lambda^{\tau \tau^\prime} Y^{\tau^\prime \tilde\sigma_i}.
\end{align}
Now the local reduced density matrix can be written as
\begin{align}
(\rho_i)^{\tilde\sigma^\prime_i \tilde\sigma_i} = \Lambda^{\tau \tau^\prime} Y^{\tau^\prime \tilde\sigma^\prime_i} Y^{\dagger \tilde\sigma_i \tau^{\prime\prime}} \Lambda^{\tau^{\prime\prime}\tau}.
\end{align}
We then perform a DMRG optimization step on $\tilde R^{\tau \sigma_i} = \Lambda^{\tau \tau^\prime} Y^{\tau^\prime \tilde\sigma_i} R^{\tilde\sigma_{i} \sigma_{i}}$ using a Lanczos optimization scheme. This step optimizes the local basis for the current MPS.

The next step is to shift the focus back to the local site tensor [Fig.~\ref{fig:sketch_dmrg_lbo}(b)]. We again perform an SVD, now  on the optimized $\tilde R^{\tau \sigma_i}$ tensor and in the process truncate the new optimized basis to the desired size:
\begin{align}
\tilde R^{\tau \sigma_i} = \tilde X^{\tau \tilde\sigma^\prime_i} \tilde\Lambda^{\tilde\sigma^\prime_i \tilde\sigma_i} \tilde Y^{\tilde\sigma_i \sigma_i}.
\end{align}
We set the $\tilde Y^{\tilde\sigma_i \sigma_i}$ as our new local basis transformation matrix and our new site tensor is
\begin{align}
\tilde M^{\tilde\sigma_{i}}_{a_{i-1} a_{i}} = X^\tau_{a_{i-1} a_i} \tilde X^{\tau \tilde\sigma^\prime_i} \tilde\Lambda^{\tilde\sigma^\prime_i \tilde\sigma_i}.
\end{align}
The third step is  to perform a single-site DMRG optimization on the new $\tilde M^{\tilde\sigma_{i}}_{a_{i-1} a_{i}}$ tensor using the local Hamiltonian in matrix-product operator form. The local Hamiltonian can be transformed into the optimized basis using the updated tensor $R^{\tilde\sigma_i \sigma_i}$ [Fig.~\ref{fig:sketch_dmrg_lbo}(c)].

In principle, these three steps can be repeated several times until no further improvements can be detected.
However, in the implementation used for this work we fix the number of iterations to just one or two.

When we shift the focus to the next site we perform a subspace expansion as explained in Ref.~\onlinecite{Hubig-PRB-2015} to avoid getting stuck in local minima in the energy landscape.

\begin{figure}[!tb]
\includegraphics[width=\columnwidth]{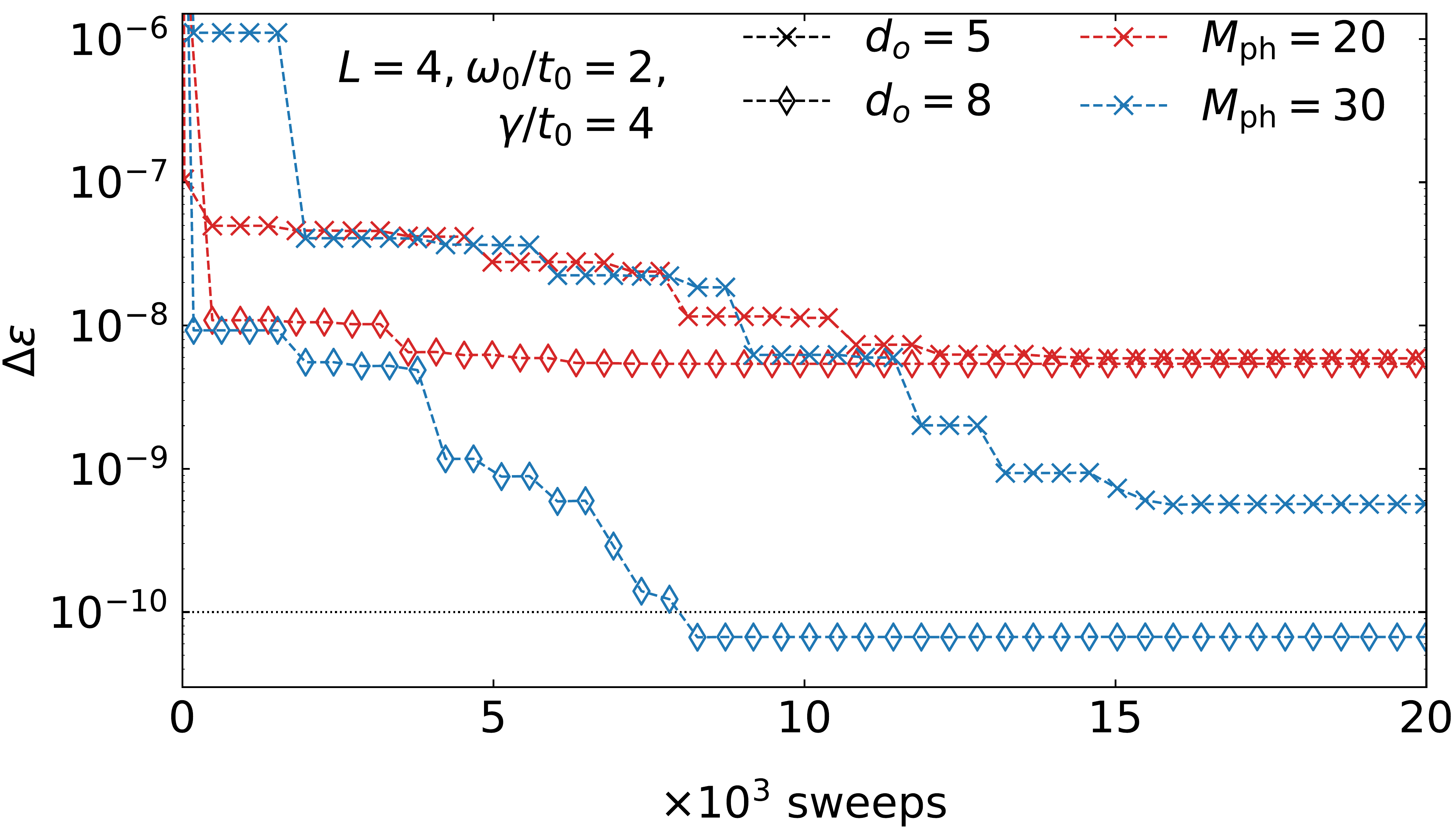}
\caption{\label{fig:gs_convergence_L4g2}Relative error $\Delta \epsilon$ of the ground-state energy obtained with the DMRG3S+LBO algorithm calculated for $L = 4, \omega_0/t_0 = 2$, and $\gamma/t_0 = 4$
(crosses: data calculated with $d_o = 5$, diamonds: $d_o = 8$). Red symbols were calculated with $M_{\rm ph} = 20$ and blue symbols with $M_{\rm ph} = 30$. The discarded weight for the bond dimension is $10^{-10}$. The thin black dotted line marks $\Delta\epsilon = 10^{-10}$.}
\end{figure}

The first two steps of the local optimization described above can be combined with any DMRG algorithm.
However, using single-site DMRG is especially beneficial here since such an algorithm scales better with the local (optimal) basis dimension.
For example, for spinless fermions or the Fermi-Hubbard model, the local dimension is $d=2$ or $d=4$, respectively.
Utilizing symmetry sectors (e.g., particle-number conservation), the effective local dimension of every symmetry block can be reduced down to $d_{\rm eff} = 1$.
As a consequence, the local Hilbert-space dimension is more or less irrelevant for the performance of the algorithm (the runtime scales at most linearly with the number of symmetry blocks).
Therefore, single-site DMRG algorithms have no major performance benefit over a two-site DMRG algorithm.
However, for systems such as the Holstein model, where some degrees of freedom are not conserved (i.e., the number of phonons), the scaling of the algorithm with the local dimension becomes substantial.
Away from the atomic limit, $t_0 \neq 0$, the local dimension is $d_o > 1$ in the different symmetry blocks and, as a consequence, an efficient single-site DMRG algorithm is desirable.

In the implementation used for this work, we utilize the fermion number conservation of the Hamiltonian Eq.~\eqref{eq:hol-ham}.
This means that the local basis transformation tensors $R$ consist of two symmetry blocks. In our algorithm, we fix a maximal dimension $d_o$ of the blocks.
In the truncation process [Fig.~\ref{fig:sketch_dmrg_lbo}(b)] we take the singular values of both blocks of $\tilde\Lambda$, sort them by size, and then start filling the blocks starting with the largest singular value.
We stop as soon as one of the blocks has reached the maximal dimension $d_o$.

In order to test the validity of our approach, we compare the DMRG3S+LBO results with Lanczos diagonalization that produces numerically exact results \cite{Prelovsek-2013}. As already mentioned in the introduction, the unbounded Hilbert space of the bosonic phonon degrees of freedom requires an {\it ad hoc} cutoff in order to be feasible for exact wave-function based methods. In Fig.~\ref{fig:gs_convergence_L4g2}, we show the relative error of the ground-state energy, i.e.,
\begin{align}
\Delta\varepsilon = \frac{E_{\rm DMRG} - E_{\rm Lz}}{|E_{\rm Lz}|}\,,
\end{align}
where $E_{\rm DMRG}$[$E_{\rm Lz}$] stands for the ground-state energy obtained with DMRG3S+LBO [Lanczos]. In order to compare with Lanczos  diagonalization, we investigate a small system of $L = 4$ in the CDW phase ($\omega_0/t_0 = 2$ and $\gamma/t_0 = 4$). In the Lanczos approach, we use $M_{\rm Lz}=400$ Lanczos steps and $M_{\rm ph}=30$ phonons per site, which yields a Hilbert space of $\mbox{dim}(\mathcal{H})\simeq 5\cdot10^6$ at half filling. In the DMRG3S+LBO ground-state search, we fix the discarded weight in the bond dimension to $10^{-10}$. We present $\Delta\varepsilon$  for different maximal phonon numbers per site $M_{\rm ph}$ as different colors and different maximal numbers of optimal modes per fermion sector $d_o$ as different symbols in Fig.~\ref{fig:gs_convergence_L4g2}.

It is evident from the presented results that one needs to converge in both the number of optimal modes $d_o$ and the maximal local phonon number $M_{\rm ph}$ to get an accurate result.
One can see that a maximum number of optimal modes of $d_o = 5$ or a maximum local phonon number of $M_{\rm ph} = 20$ is not enough to get an energy with an error of the same order as the discarded weight $10^{-10}$ (thin black dotted line in Fig.~\ref{fig:gs_convergence_L4g2}).
To converge the energy difference $\Delta\varepsilon$ to the same order of magnitude as the discarded weight, a minimum number of phonons per site of $M_{\rm ph} = 30$ and a minimum number of optimal modes of $d_o = 8$ is required.
Comparing the convergence behavior for different parameter sets (not shown here), we observe that our DMRG3S+LBO method is especially well suited for the region where $t_0 \sim \omega_0$. 

As mentioned above, we use a subspace expansion to avoid local minima in the energy landscape when converging to the ground state \cite{Hubig-PRB-2015}.
Within this scheme, a mixing factor is introduced that controls the MPS-basis enrichment process.
As pointed out in Ref.~\onlinecite{Hubig-PRB-2015}, it is a delicate task to choose this mixing factor in such a way that one avoids local minima while still converging in energy.
This seems to be especially hard when working with a fixed discarded weight in the bond dimension.
To check convergence of the algorithm, it is advisable to not only monitor the ground-state energy during the runs but also the variance of the energy $\sigma_E^2 = \langle \psi | H^2 | \psi \rangle - \langle \psi | H | \psi \rangle^2$.
The variance can be taken as a measure of how close a given state is to an eigenstate of the Hamiltonian.

In the present work, the DMRG3S+LBO algorithm will be used for comparatively small system sizes since these are constrained by what can be handled with the following time-evolution method.
However, in the Appendix, we show that our findings are robust against finite-size effects.
A more extensive discussion of the DMRG3S+LBO method and a benchmark against other state-of-the-art DMRG methods for 
electron-phonon systems such as the pseudosite method \cite{Jeckelmann-PRB-1998} and the method introduced in Ref.~\cite{Kloss-PRL-2019} will be presented elsewhere. 

\subsection{TEBD with local basis optimization}
\label{sec:tebd+lbo}
As discussed in the previous section, rotating the local basis into an optimized basis can be beneficial for MPS-based numerical methods.
For the time evolution used in this work, we therefore employ the same strategy.
We use the time-evolving block decimation (TEBD) algorithm pioneered by Vidal \cite{Vidal-PRL-2004,Vidal-PRL-2003} and combine it with local basis optimization \cite{Zhang-PRL-1998}.
The algorithm used here is based on Ref.~\cite{Brockt-PRB-2015}, where single-electron problems are studied, and applies this method to finite electron densities.
In the following, we will outline the different steps in this time-evolution approach.

\begin{figure}[!tb]
\includegraphics[width=\columnwidth]{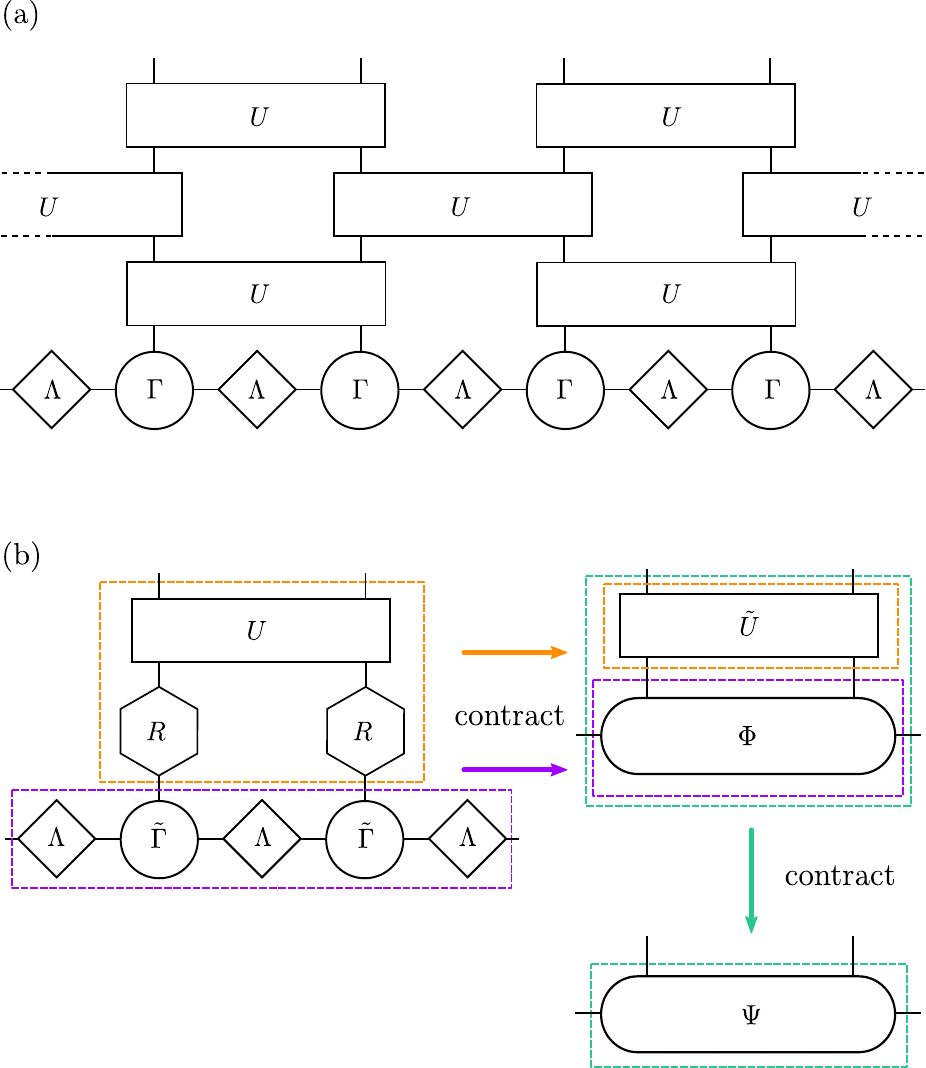}
\caption{\label{fig:sketch_tebd_lbo} (a) General structure of a TEBD algorithm \cite{Schollwoeck-AoP-2011,Vidal-PRL-2004}. (b) Different steps in the application of a single local time-evolution operator in the TEBD-LBO algorithm \cite{Brockt-PRB-2015}.}
\end{figure}

The time-evolving block decimation relies on the Trotter decomposition of the Hamiltonian.
Consider the Hamiltonian $H_{\rm NN}$ of a one-dimensional lattice system with at most a nearest-neighbor interaction. Then, $H_{\rm NN}$ can be split into two sums:
\begin{align}
H_{\rm NN} = \sum_{l = 1}^L h_l = \sum_{l \ \rm odd} h_l + \sum_{l \ \rm even} h_l =  H_{\rm odd} + H_{\rm even}\,,
\end{align} 
where all local summands $h_l$ in $H_{\rm even}$ and $H_{\rm odd}$ commute with each other. The corresponding time-evolution operator can be written in a second-order Trotter decomposition as
\begin{align}
e^{-\im H_{\rm NN} \delta t} = \ &e^{-\im H_{\rm odd} \delta t/2} e^{-\im H_{\rm even} \delta t} e^{-\im H_{\rm odd} \delta t/2} \nonumber \\ 
&+ \mathcal{O}((\delta t)^3) \nonumber \\
= \ &\prod_{l \ \rm odd} e^{-\im h_{l} \delta t/2} \prod_{l \ \rm even} e^{-\im h_{l} \delta t} \prod_{l \ \rm odd} e^{-\im h_{l} \delta t/2} \nonumber \\
&+ \mathcal{O}((\delta t)^3)\,.
\end{align}
The individual local time-evolution operators $U_l = e^{-\im h_{l} \delta t}$ only act on two adjacent sites. In the MPS algorithm, these $U_l$ operators take the form of gates that are applied to the MPS [Fig.~\ref{fig:sketch_tebd_lbo}(a)].

Consider a generic MPS in Vidal's notation \cite{Vidal-PRL-2003} where on every site, there is an additional basis transformation tensor $R$ as in Eq.~\eqref{eq:mps-lbo}:
\begin{align}
\left| \psi \right\rangle = \sum_{\{ \sigma_l \}} \tilde \Gamma^{\tilde\sigma_1}_{a_0 a_1} R^{\tilde\sigma_1 \sigma_1} \Lambda^{[1]}_{a_1 a^\prime_1} \tilde \Gamma^{\tilde\sigma_2}_{a^\prime_1 a_2} R^{\tilde\sigma_2 \sigma_2} \Lambda^{[2]}_{a_2 a^\prime_2} ... \nonumber \\ 
\Lambda^{[L-1]}_{a_{L-1} a^\prime_{L-1}} \Gamma^{\tilde\sigma_L}_{a^\prime_{L-1} a_L} R^{\tilde\sigma_L \sigma_L} \left| \sigma_1 ... \sigma_L \right\rangle \,.
\end{align}

The first step in the time evolution is to contract the local basis transformation from one side to the local time-evolution operators $U_l$ while the other side stays in the original basis [Fig.~\ref{fig:sketch_tebd_lbo}(b)]:
\begin{align}
R^{\tilde\sigma_l \sigma_l} R^{\tilde\sigma_{l+1} \sigma_{l+1}} U_l^{\sigma_l \sigma_{l+1} \sigma^\prime_l \sigma^\prime_{l+1}} = \tilde U_l^{\tilde\sigma_l \tilde\sigma_{l+1} \sigma^\prime_l \sigma^\prime_{l+1}}\,.
\end{align}
With this modified time-evolution operator $\tilde U_l$ we act on the bond tensor $\Phi$ [Fig.~\ref{fig:sketch_tebd_lbo}(b)]:
\begin{align}
\Phi_{a_{l-1} a^\prime_{l+1}}^{\tilde\sigma_l \tilde\sigma_{l+1}} &= \Lambda^{[l-1]}_{a_{l-1} a^\prime_{l-1}} \Gamma^{\tilde\sigma_l}_{a^\prime_{l-1} a_l} \Lambda^{[l]}_{a_l a^\prime_l} \Gamma^{\tilde\sigma_{l+1}}_{a^\prime_{l} a_{l+1}} \Lambda^{[l+1]}_{a_{l+1} a^\prime_{l+1}} \\
\Psi_{a_{l-1} a^\prime_{l+1}}^{\sigma^\prime_l \sigma^\prime_{l+1}} &= \Phi_{a_{l-1} a^\prime_{l+1}}^{\tilde\sigma_l \tilde\sigma_{l+1}} \tilde U_l^{\tilde\sigma_l \tilde\sigma_{l+1} \sigma^\prime_l \sigma^\prime_{l+1}}\,.
\end{align}
Note that the updated bond tensor $\Psi$ is now in the original basis.
This is important to ensure that during the time evolution, the full local Hilbert space can be explored and also the optimal modes can change from before to after the application of the time-evolution operator. Next, we transform the time-evolved bond tensor $\Psi$ to the optimized basis. For that we calculate the local reduced density matrix on the sites $l$ and $l+1$:
\begin{align}
\rho^{\sigma^\prime_l \sigma^{\prime \prime}_l} &= \Psi_{a_{l-1} a^\prime_{l+1}}^{\sigma^\prime_l \sigma^\prime_{l+1}} \Psi_{a^\prime_{l+1} a_{l-1} }^{\dagger \sigma^\prime_{l+1} \sigma^{\prime\prime}_l} \\
\rho^{\sigma^\prime_{l+1} \sigma^{\prime \prime}_{l+1}} &= \Psi_{a_{l-1} a^\prime_{l+1}}^{\sigma^\prime_l \sigma^\prime_{l+1}} \Psi_{a^\prime_{l+1} a_{l-1} }^{\dagger \sigma^{\prime\prime}_{l+1} \sigma^\prime_l}\,.
\end{align}
Next, we diagonalize the local reduced density matrices to obtain the local basis transformation matrices $U^\dagger$.
Each of them can then be truncated to the desired optimal dimension $d_o$ to obtain the basis transformation matrices $R_l$ and $R_{l+1}$.
For the time evolution, we actually define a local discarded weight $\Delta_{\rm loc}$ which is the maximum weight that is discarded from the spectrum of the local density matrix.
We keep this local discarded weight fixed rather than the optimal dimension throughout one simulation.
By applying the inverse of $R$ on the new bond tensor $\Psi$, we get the bond tensor in the optimal basis $\tilde\Psi$.
We then go back to the original Vidal notation by performing an SVD of the $\tilde\Psi = U S V^\dagger$ and contracting the inverse of $\Lambda^{[l-1]}$ from the left to $U$ and the inverse of $\Lambda^{[l+1]}$ from the right to $V^\dagger$ to obtain $\Gamma^{\tilde\sigma_l}$, $\Lambda^{[l]}$ and $\Gamma^{\tilde\sigma_{l+1}}$.

In a conventional time-dependent DMRG method, one has the discarded weight in the bond dimension $\Delta_{\rm tr}$ and the time-step size $\delta t$ as simulation parameters. In the TEBD with local basis optimization (TEBD-LBO) algorithm, one additionally gets the maximal local phonon number $M_{\rm ph}$ and the local discarded weight $\Delta_{\rm loc}$ as simulation parameters.
For all of the results presented in Sec.~\ref{sec:nonequ}, we made sure that the total error originating from $\delta t$, $M_{\rm ph}$, $\Delta_{\rm tr}$, and $\Delta_{\rm loc}$ is smaller than the symbol size (as a consequence, the error bars are omitted in all figures). This is achieved by setting $\Delta_{\rm tr}, \Delta_{\rm loc} \leq 10^{-7}$ throughout the paper and choosing $\delta t\,t_0 \leq 0.05$.
As opposed to the method used by Brockt {\it et al.} \cite{Brockt-PRB-2015}, where the maximum number of phonons per site can grow during the time evolution, we work with a fixed maximal phonon number $M_{\rm ph}$ per site.

\begin{figure}[!tb]
\includegraphics[width=\columnwidth]{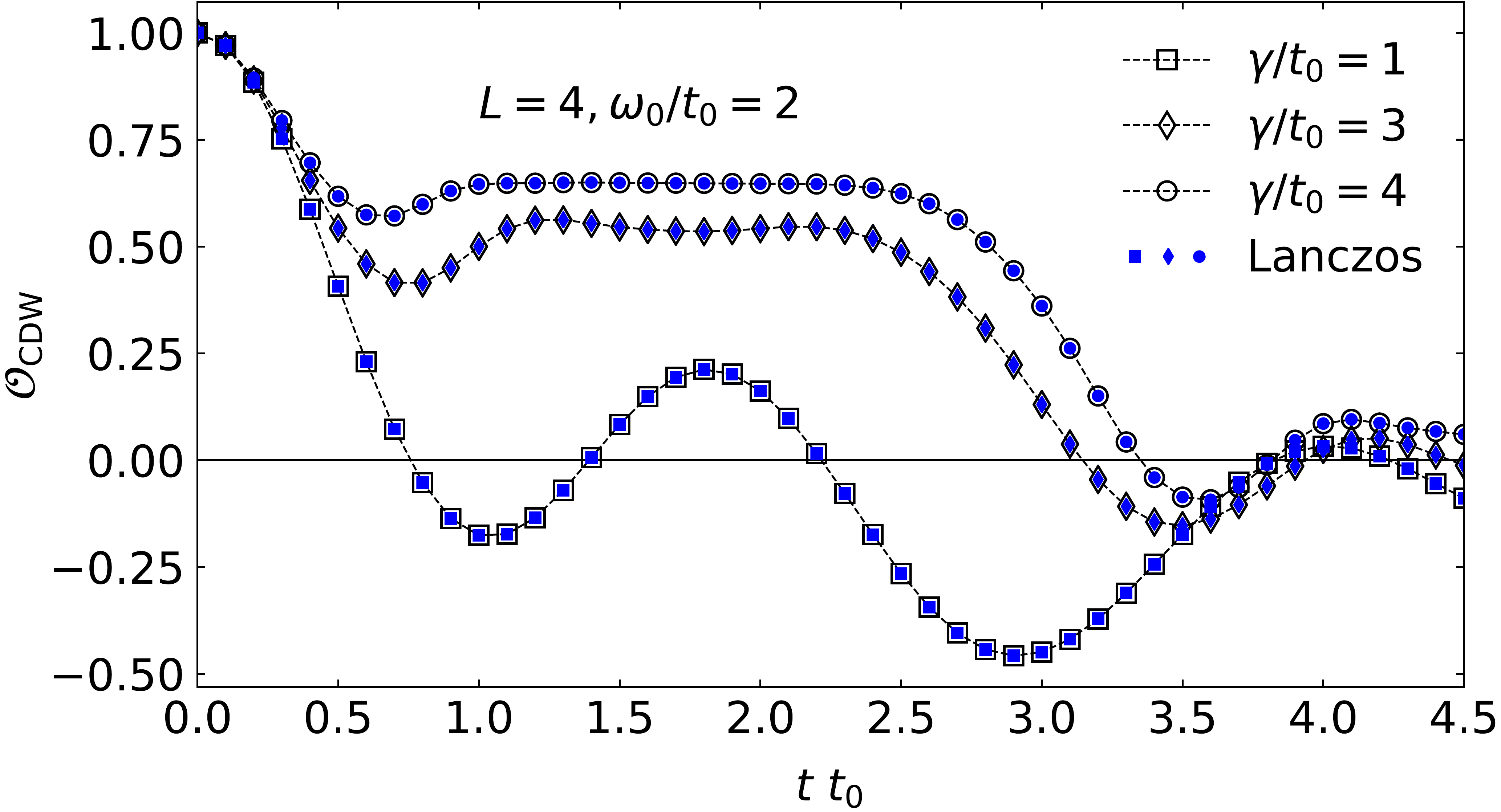}
\caption{\label{fig:bare_Ocdw_Lanczos}Comparison between TEBD-LBO data (open black symbols) and Lanczos time-evolution data (small blue symbols) of the decay of the charge-density-wave order parameter $\mathcal{O}_{\rm CDW}$ starting from the bare CDW state $|\mathrm{BCDW}\rangle$ Eq.~\eqref{eq:BCDW}. Calculations are done for system size $L=4$, phonon frequency $\omega_0/t_0 = 2$, and different coupling strengths $\gamma/t_0 = 1, 3, 4$ (squares, diamonds, and circles, respectively). In the TEBD-LBO time evolution, we use a local phonon cutoff $M_{\rm ph} = 10, 30, 40$, respectively. The local discarded weight is set to $\Delta_{\rm loc} = 10^{-8}$. For clarity, we only show every fourth data point that was computed in TEBD-LBO and every twentieth data point from the Lanczos time evolution.}
\end{figure}

Let us now test the accuracy of the TEBD-LBO algorithm. In Fig.~\ref{fig:bare_Ocdw_Lanczos}, we present the decay of $\mathcal{O}_{\rm CDW}$ starting from a CDW state without phonons, i.e., $|\psi(\tau=0)\rangle=|0101\rangle_{\rm el}|\emptyset\rangle_{\rm ph}$ (with $|\emptyset\rangle_{\rm ph}$ as the vacuum state of the phonons; see also Sec.~\ref{sec:bare} for details) as calculated with TEBD-LBO and Lanczos time evolution for system size $L=4$. The time evolution within the latter is carried out with a time step of $\delta t\,t_0=10^{-2}$ and $M_{\rm Lz}=20$ Lanczos steps.
It is evident from the presented data that, similarly to DMRG3S+LBO, the TEBD-LBO algorithm perfectly reproduces the Lanczos data for all considered values of the coupling strength $\gamma$. Furthermore, we have checked (not shown) that the time evolution from other initial states (discussed in Sec.~\ref{sec:dressed} and Sec.~\ref{sec:quench}) is in full agreement with the Lanczos results.

\section{Results for the real time evolution}\label{sec:nonequ}

In this section, we present the main findings of our work: a study of the melting of CDW order during the time evolution from initial product states (see Secs.~\ref{sec:bare} and \ref{sec:dressed}) and after quenches from correlated ground states (see Sec.~\ref{sec:quench}).
In order to get a nonzero value of the CDW order parameter $\mathcal{O}_{\rm CDW}$ in the correlated ground state, we work with an odd number of sites $L$.
As a consequence, we are not exactly at half filling but rather $N = (L-1)/2$.
For consistency, we also use odd system sizes $L$ in Secs.~\ref{sec:bare} and \ref{sec:dressed}.

\begin{figure}[!tb]
\includegraphics[width=0.8\columnwidth]{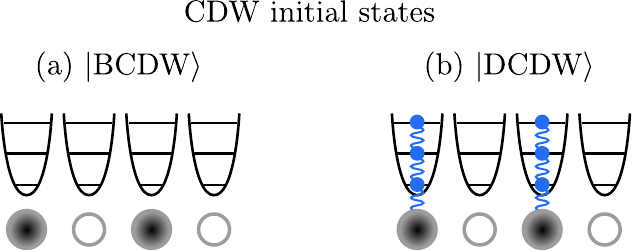}
\caption{\label{fig:init_sketch} Sketch of the initial states: (a) $|\mathrm{BCDW}\rangle$ Eq.~\eqref{eq:BCDW} and (b) $|\mathrm{DCDW} \rangle$ Eq.~\eqref{eq:DCDW}.}
\end{figure}

\subsection{\label{sec:bare}Bare CDW melting}
As a first example of charge-density-wave melting in the Holstein model we prepare the system in a product state where every second site is occupied by a fermion and no phonons are present in the system:
\begin{align}
| \mathrm{BCDW} \rangle = \left[ \prod_{l=1}^{(L-1)/2} c_{2l}^\dagger \right] | \emptyset \rangle _{\rm el}  | \emptyset \rangle _{\rm ph}\,. \label{eq:BCDW}
\end{align}
$| \emptyset \rangle _{\rm el[ph]}$ is the vacuum state of the electrons [phonons]. We call this state a bare charge-density wave (BCDW). The structure of the state in real space is sketched in Fig.~\ref{fig:init_sketch}(a). Next, we time-evolve this state
\begin{align}
| \mathrm{BCDW(t)} \rangle = e^{- \im H_{\rm Hol} t} | \mathrm{BCDW} \rangle
\end{align}
with the Hamiltonian Eq.~\eqref{eq:hol-ham} of the Holstein model for different parameter sets.

\begin{figure}[!tb]
\includegraphics[width=\columnwidth]{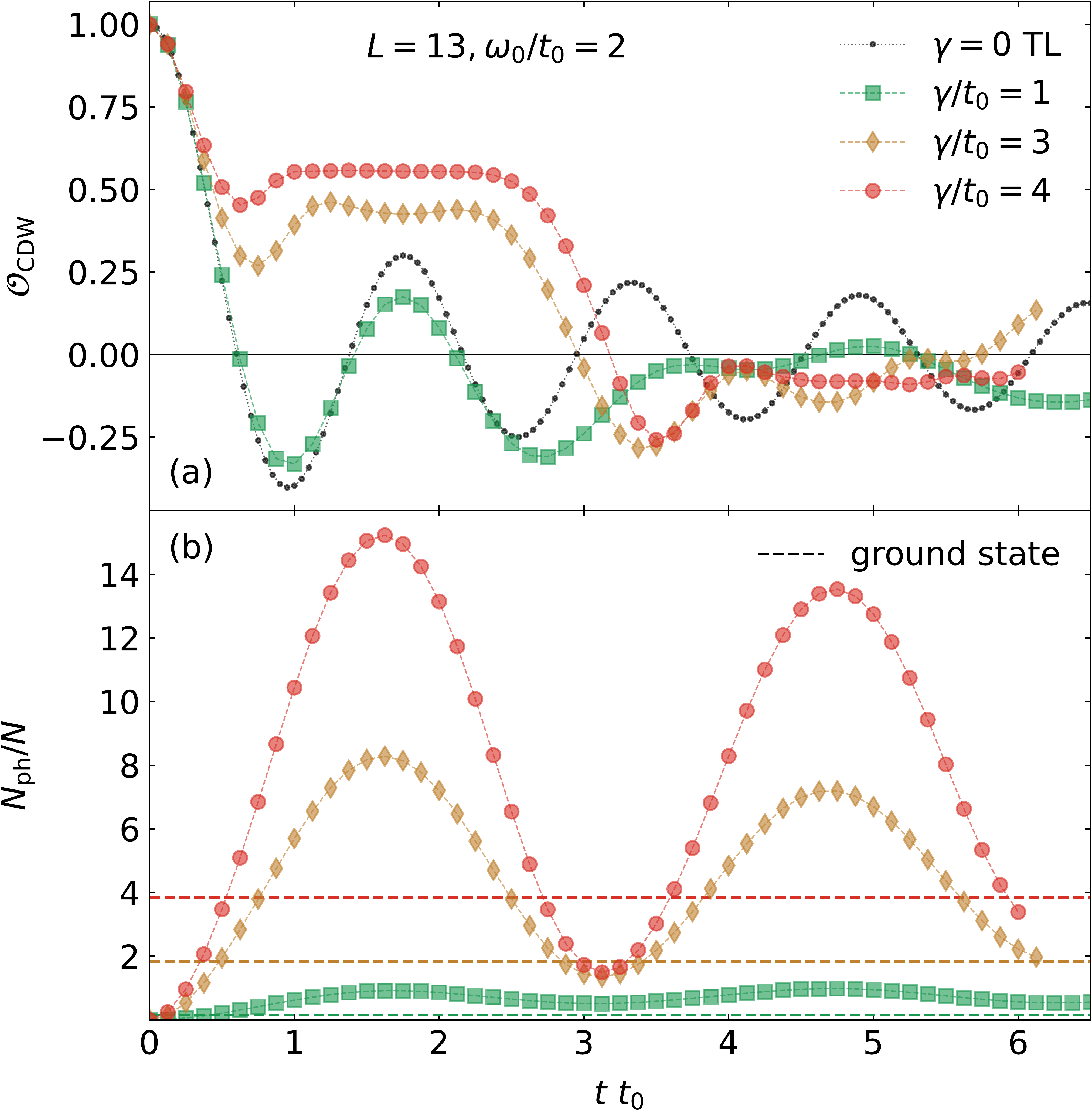}
\caption{\label{fig:bare-Ocdw-t05} Time evolution of (a) the decay of the charge-density-wave order parameter $\mathcal{O}_{\rm CDW}$ and (b) the phonon number per fermion $N_{\rm ph}/N$ when starting from the bare CDW state $|\mathrm{BCDW}\rangle$ Eq.~\eqref{eq:BCDW}. The small black dots in panel (a) are exact analytical results for $\gamma = 0$ in the thermodynamic limit \cite{Barmettler-PRL-2009}. The dashed horizontal lines in panel (b) represent the phonon number in the ground state at the respective parameters. Simulations are performed for $L=13$, $\omega_0/t_0 = 2$ and different coupling strengths $\gamma/t_0 = 1, 3, 4$. In the time evolution, we use a local phonon cutoff $M_{\rm ph} = 10, 30, 40$, respectively. The local discarded weight is set to $\Delta_{\rm loc} = 10^{-8}$. For clarity, we only show every fifth data point that was computed.}
\end{figure}

In Fig.~\ref{fig:bare-Ocdw-t05}(a), we plot the time evolution of the CDW order parameter $\mathcal{O}_{\rm CDW}$ when starting from the bare charge-density-wave state for $L=13$, $\omega_0/t_0 = 2$, and coupling strengths $\gamma/t_0 = 1, 3, 4$.
These parameter sets correspond to the TLL phase, the transition region, and the CDW phase, respectively \cite{Creffield-EPJB-2005,Bursill-PRL-1998}.
As expected for small values of  $\gamma/t_0 = 1$, the order parameter decays toward zero and oscillates around this value with an amplitude that slowly dies out. In the same figure, we compare the behavior at $\gamma/t_0 = 1$ to the behavior at $\gamma = 0$ for which the time evolution of $\mathcal{O}_{\rm CDW}$ can be calculated analytically in the thermodynamic limit, i.e., $\mathcal{O}_{\mathrm{CDW},\gamma = 0}(t) = J_0(4tt_0)$, where $J_0$ is the zeroth-order Bessel function of the first kind (see, e.g.,~\cite{Barmettler-PRL-2009}).
From this comparison, it is evident that the frequency of the oscillations is controlled by the hopping parameter $t_0$.
However, in contrast to the case of $\gamma = 0$ where the oscillations are very long lived and the amplitude decays algebraically, at $\gamma/t_0 = 1$, the amplitude of the oscillations is strongly damped.

On the contrary, for the large coupling strength $\gamma/t_0 = 4$, the order parameter, after an initial fast drop, is temporarily stuck at $\mathcal{O}_{\rm CDW} \approx 0.6$ between $ t t_0 \approx 1$ and $t t_0 \approx 2.5$ before it eventually decays toward zero.
Such a plateau is also clearly visible at coupling strength $\gamma/t_0 = 3$.
This behavior of the order parameter at strong coupling can be understood as follows.
When starting from the bare charge-density-wave state the fermions are free to move around.
By tunneling into empty sites, the fermions reduce the order imprinted in the initial state.
However, at large couplings the fermions have a strong tendency to form heavy polarons; i.e., many phonons are created as can be seen in Fig.~\ref{fig:bare-Ocdw-t05}(b) where we plot the time evolution of the number of phonons per fermion in the system $N_{\rm ph}/N = (1/N) \sum_{l=1}^L \langle b_l^\dagger b_l^{\vphantom{\dagger}} \rangle$.
These phonons surrounding the fermions drastically change their effective mass and they form heavy and therefore less mobile polarons.
As their movement is impeded, the order parameter does not change for a time span of $\approx 1.5/t_0$.
This self-trapping effect is, however, only temporary. The system coherently oscillates between a state with a large and a small number of phonons and the order parameter decays further as soon as the phonons are reemitted, allowing the electron to move again into empty sites. The phonon oscillation period can clearly be seen in the time evolution of the phonon density in the system shown in Fig.~\ref{fig:bare-Ocdw-t05}(b).
The phonon number $N_{\rm ph}/N$ oscillates with a period of $2 \pi /\omega_0$ and the length of the plateaus in $\mathcal{O}_{\rm CDW}$ at $\gamma/t_0 = 3,4$ is controlled by this phonon oscillation period.

\begin{figure}[!tb]
\includegraphics[width=\columnwidth]{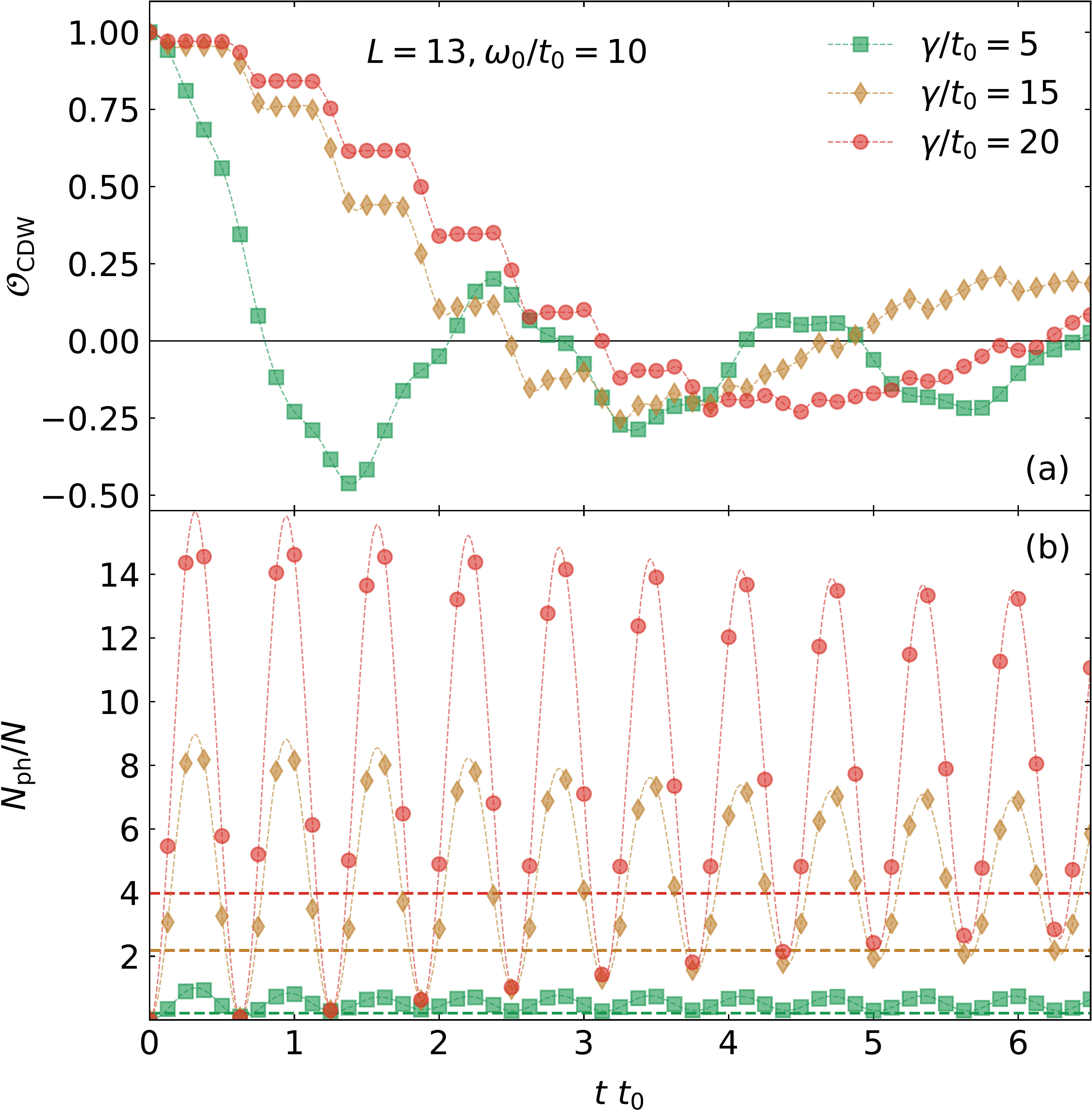}
\caption{\label{fig:bare-Ocdw-t01} Time evolution of (a) the decay of the charge-density-wave order parameter $\mathcal{O}_{\rm CDW}$ and (b) the phonon number per fermion $N_{\rm ph}/N$ when starting from the bare CDW state $|\mathrm{BCDW}\rangle$ Eq.~\eqref{eq:BCDW}. Simulations are performed for  $L=13$,  $\omega_0/t_0 = 10$ and different coupling strengths $\gamma/t_0 = 5, 15, 20$. In the time evolution, we use a local phonon cutoff $M_{\rm ph} = 10, 30, 40$, respectively, and the local discarded weight is set to $\Delta_{\rm loc} = 10^{-8}$. For clarity, we only show every twentyfifth data point that was computed.}
\end{figure}

If one further increases the phonon frequency $\omega_0/t_0$, several plateaus can be observed before the order parameter $\mathcal{O}_{\rm CDW}$ relaxes toward zero.
Such a behavior can be seen in Fig.~\ref{fig:bare-Ocdw-t01}(a) where we plot the time evolution of $\mathcal{O}_{\rm CDW}$ for the same initial state $|\mathrm{BCDW}\rangle$ but for $\omega_0/t_0 = 10$.
The steplike structure in the decay of the order parameter is evident in the data for $\gamma/t_0 = 15,20$ and the length of the plateaus coincides well with the phonon oscillation period $2\pi/\omega_0$ [see Fig.~\ref{fig:bare-Ocdw-t01}(b) for the time dependence of the phonon density $N_{\rm ph}/N$ in the system].
Similar to the case at $\omega_0/t_0 = 2$, for the weaker coupling $\gamma/t_0 = 5$, we observe a decay of the order parameter toward zero with damped oscillations with a frequency controlled by the hopping parameter $t_0$. These oscillations are superimposed with oscillations that have a frequency controlled by the phonon frequency $\omega_0$.

\begin{figure}[!tb]
\includegraphics[width=\columnwidth]{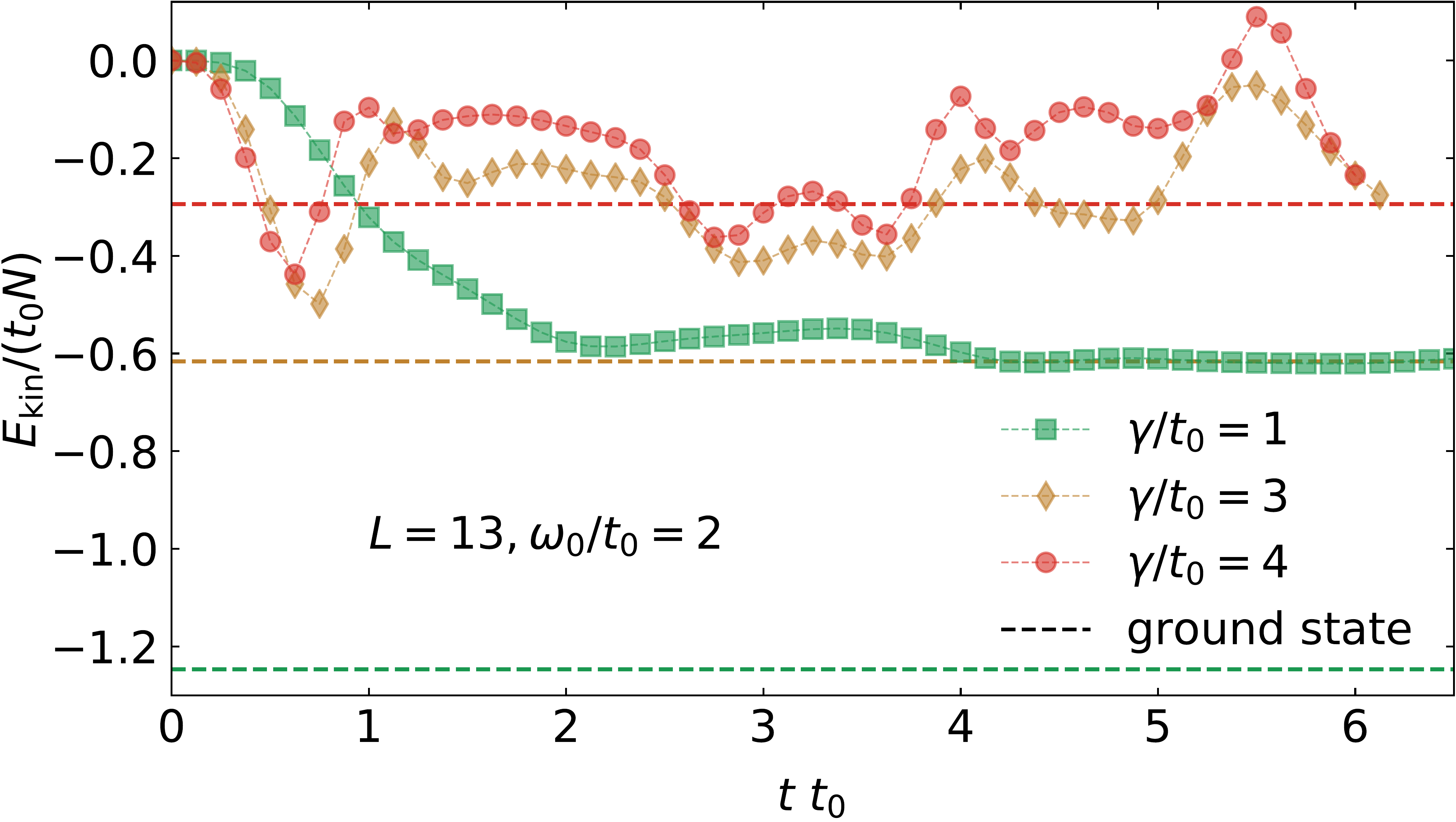}
\caption{\label{fig:bare-Ekin-t05} Time evolution of the kinetic energy per fermion $E_{\rm kin}/N$ when starting from the bare CDW state $|\mathrm{BCDW}\rangle$ Eq.~\eqref{eq:BCDW}. The dashed horizontal lines represent the kinetic energy in the ground state at the respective parameters. Simulations are performed for $L=13$, $\omega_0/t_0 = 2$ and different coupling strengths $\gamma/t_0 = 1, 3, 4$. In the time evolution, we use a local phonon cutoff $M_{\rm ph} = 10, 30, 40$, respectively. The local discarded weight is set to $\Delta_{\rm loc} = 10^{-8}$. For clarity, we only show every fifth data point that was computed.}
\end{figure}

In Fig.~\ref{fig:bare-Ekin-t05}, we plot the kinetic energy $E_{\rm kin} = \langle H_{\rm kin} \rangle$ of the fermions as a function of time for $L = 13$, $\omega_0/t_0 = 2$ and different coupling strengths $\gamma/t_0 = 1, 3, 4$ (i.e., the same parameters as in Fig.~\ref{fig:bare-Ocdw-t05}).
It is not surprising that the initial drop in kinetic energy gets steeper as the coupling strength $\gamma/t_0$ is increased.
However, for longer times the energy loss from the electronic system decreases with increasing coupling strength.
This trend follows the trend of the ground-state kinetic energy plotted as dashed horizontal lines in Fig.~\ref{fig:bare-Ekin-t05}.
The electrons get more and more localized in the ground state as $\gamma/t_0$ increases and therefore, the kinetic energy grows.
Yet, we emphasize here that during the time evolution we do not drift toward the ground state since energy is conserved throughout the time evolution.
Quite on the contrary, we remain in a high-energy state.
An open question left for future work is a comparison to finite-temperature equilibrium expectation values of the same  observables.

\begin{figure}[!tb]
 \includegraphics[width=\columnwidth]{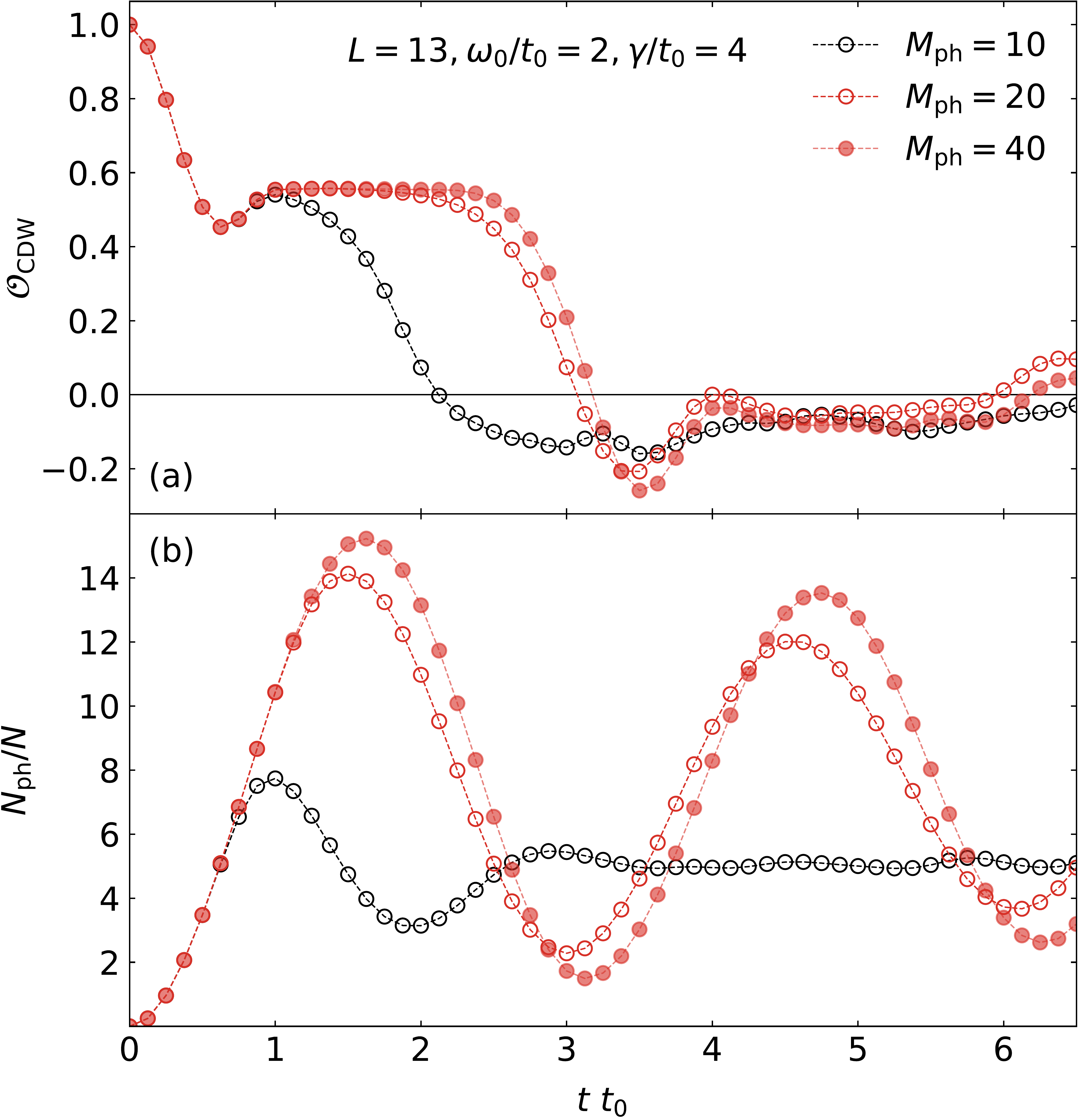}
\caption{\label{fig:compare} Time evolution of (a) the decay of the charge-density-wave order parameter $\mathcal{O}_{\rm CDW}$ and (b) the phonon number per fermion $N_{\rm ph}/N$ when starting from the bare CDW state $|\mathrm{BCDW}\rangle$ Eq.~\eqref{eq:BCDW}. Simulations are performed for  $L=13$,  $\omega_0/t_0 = 2$ and coupling strengths $\gamma/t_0 = 4$. In the time evolution, we use different local phonon cutoffs $M_{\rm ph} = 10, 20, 40$ to illustrate convergence in this parameter. The local discarded weight is set to $\Delta_{\rm loc} = 10^{-7}$. For clarity, we only show every fifth data point that was computed.}
\end{figure}

In order to illustrate the capabilities of the TEBD-LBO method, we compare such a simulation that is converged for a given local and global discarded weight
for $L=13$ sites ($\omega_0/t_0 = 2$, $\gamma/t_0 = 4$) with a resulting $d_o=12$ (and an $M_{\text{ph}}=40$) to a simulation with $M_{\text{ph}} =10$ and $M_{\text{ph}} =20$, which is shown in Fig.~\ref{fig:compare}.
Clearly, the simulation with $M_{\text{ph}} = 10$ cannot correctly produce the dynamics for $t>1/t_0$ and fails to capture the intermediate plateau formation for $1 \lesssim t t_0 \lesssim 2.5$.
The simulation with $M_{\text{ph}} = 20$ is able to capture the plateau formation but evidently is not converged.
This shows that the TEBD-LBO is not only more accurate on a quantitative level but is also capable of accessing parameter regimes that are out of reach for conventional simulations with a small $M_{\text{ph}}$ using the phonon-number basis.

\subsection{\label{sec:dressed}Dressed CDW melting}

In the second example, we start from the ground state in the atomic limit $t_0 = 0$.
As discussed in Sec.~\ref{sec:model}, the ground state takes the form Eq.~\eqref{eq:gs-lang-firsov} and we prepare it in such a way that $\mathcal{O}_{\rm CDW} = 1$.
This is done by setting the hopping parameter $t_0 = 0$ and performing an imaginary time evolution of the bare charge-density-wave state $| \mathrm{BCDW} \rangle$ to reach the ground state.
This results in the state
\begin{align}
| \mathrm{DCDW} \rangle = e^{-\frac{(L-1) \gamma^2}{4\omega_0^2}}  \left[  \prod_{l=1}^{(L-1)/2} c_{2l}^\dagger \ e^{\frac{\gamma}{\omega_0}b_{2l}^\dagger} \right] | \emptyset \rangle _{\rm el}  | \emptyset \rangle_{\rm ph}\,,  \label{eq:DCDW}
\end{align}
up to machine precision. We will refer to this state as a dressed charge-density wave (DCDW) and its structure in real space is sketched in Fig.~\ref{fig:init_sketch}(b).

\begin{figure}[!tb]
\includegraphics[width=\columnwidth]{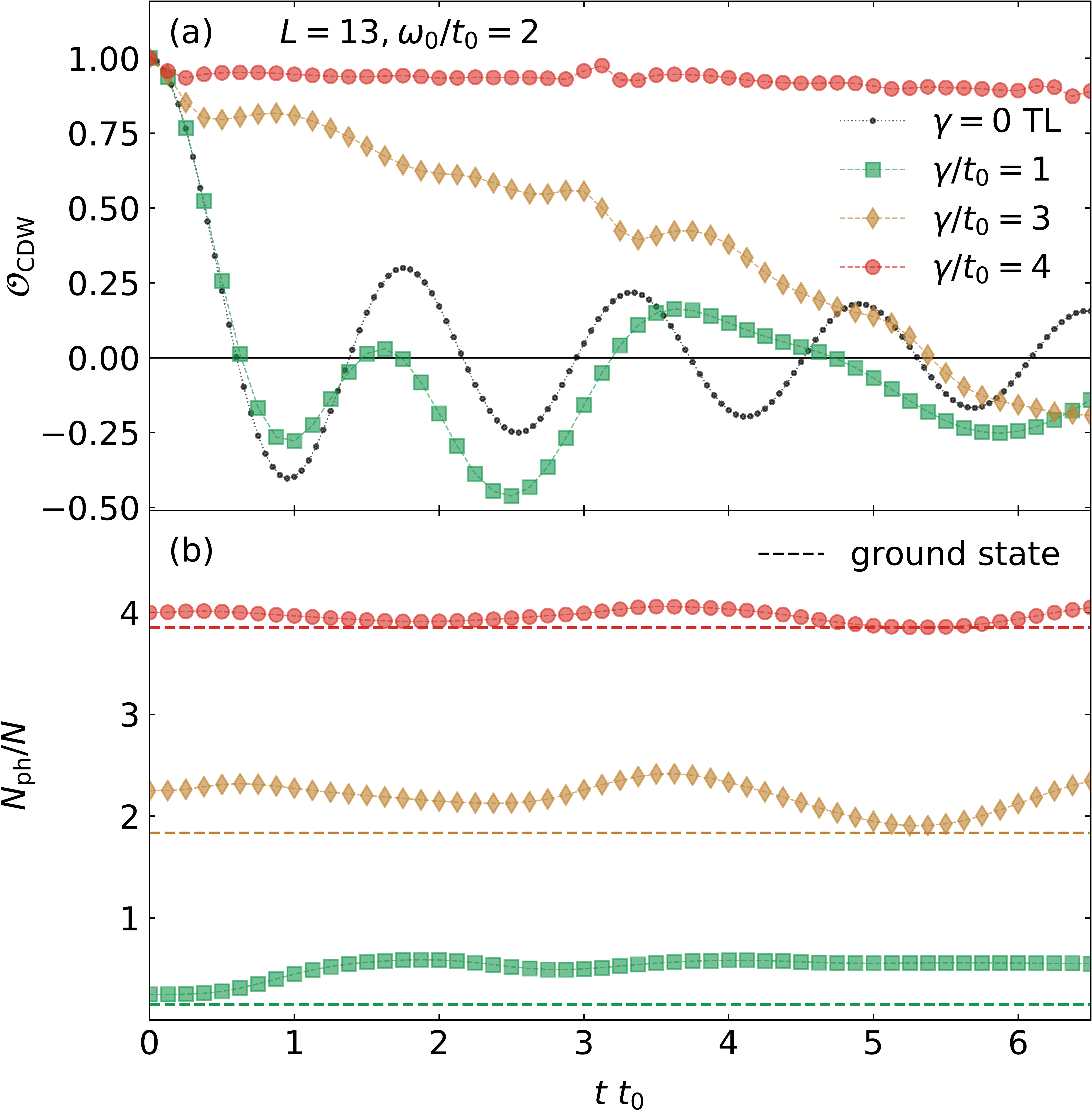}
\caption{\label{fig:dressed-Ocdw-t05} Time evolution of (a) the decay of the charge-density-wave order parameter $\mathcal{O}_{\rm CDW}$ and (b) the phonon number per fermion $N_{\rm ph}/N$ when starting from the dressed CDW state $|\mathrm{DCDW}\rangle$ Eq.~\eqref{eq:DCDW}.  The small black dots in panel (a) are exact analytical results for $\gamma = 0$ when starting from the $|\mathrm{BCDW}\rangle$ state in the thermodynamic limit \cite{Barmettler-PRL-2009}. The dashed horizontal lines in panel (b) represent the number of phonons per fermion in the ground states at the respective parameters. Simulations are performed for $L=13$, $\omega_0/t_0 = 2$ and different coupling strengths $\gamma/t_0 = 1, 3, 4$. In the time evolution, we use a local phonon cutoff $M_{\rm ph} = 20, 30, 40$, respectively. The local discarded weight is set to $\Delta_{\rm loc} = 10^{-8}$. For clarity, we only show every fifth data point that was computed.}
\end{figure}

In Fig.~\ref{fig:dressed-Ocdw-t05}(a), we plot the order parameter $\mathcal{O}_{\rm CDW}$ as a function of time when starting from the DCDW state.
We set the phonon frequency to $\omega_0/t_0 = 2$ during the time evolution and use different coupling strengths $\gamma/t_0 = 1,3,4$ (the same as for the BCDW state in Fig.~\ref{fig:bare-Ocdw-t05} and Fig.~\ref{fig:bare-Ekin-t05}).
For the strongest coupling $\gamma/t_0 = 4$, the initial state is close to the ground state and therefore, the order parameter decays very slowly.
This resemblance is also reflected in the time dependence of the phonon number plotted in Fig.~\ref{fig:dressed-Ocdw-t05}(b).
For the strong coupling $\gamma/t_0 = 4$, the phonon number barely changes over time and stays close to the value in the ground state plotted as a dashed horizontal line.
On the other hand, for the small coupling $\gamma/t_0 = 1$, the initial state is far from the ground state and, as a consequence, the order decays fast toward zero and oscillates around this value.
Again, the frequency of the oscillations is controlled by the hopping parameter $t_0$ as is evident from the comparison to the exact analytical curve at $\gamma = 0$ \cite{Barmettler-PRL-2009} [small black dots in Fig.~\ref{fig:dressed-Ocdw-t05}(a)].
Furthermore, for $\gamma/t_0 = 1$, the phonon number increases by a factor of two within $tt_0 \approx 1.5$. For the intermediate coupling of $\gamma/t_0 = 3$, the order parameter slowly and steadily decays to zero and the phonon number in the system changes moderately compared to the other two cases.

\begin{figure}[!tb]
\includegraphics[width=\columnwidth]{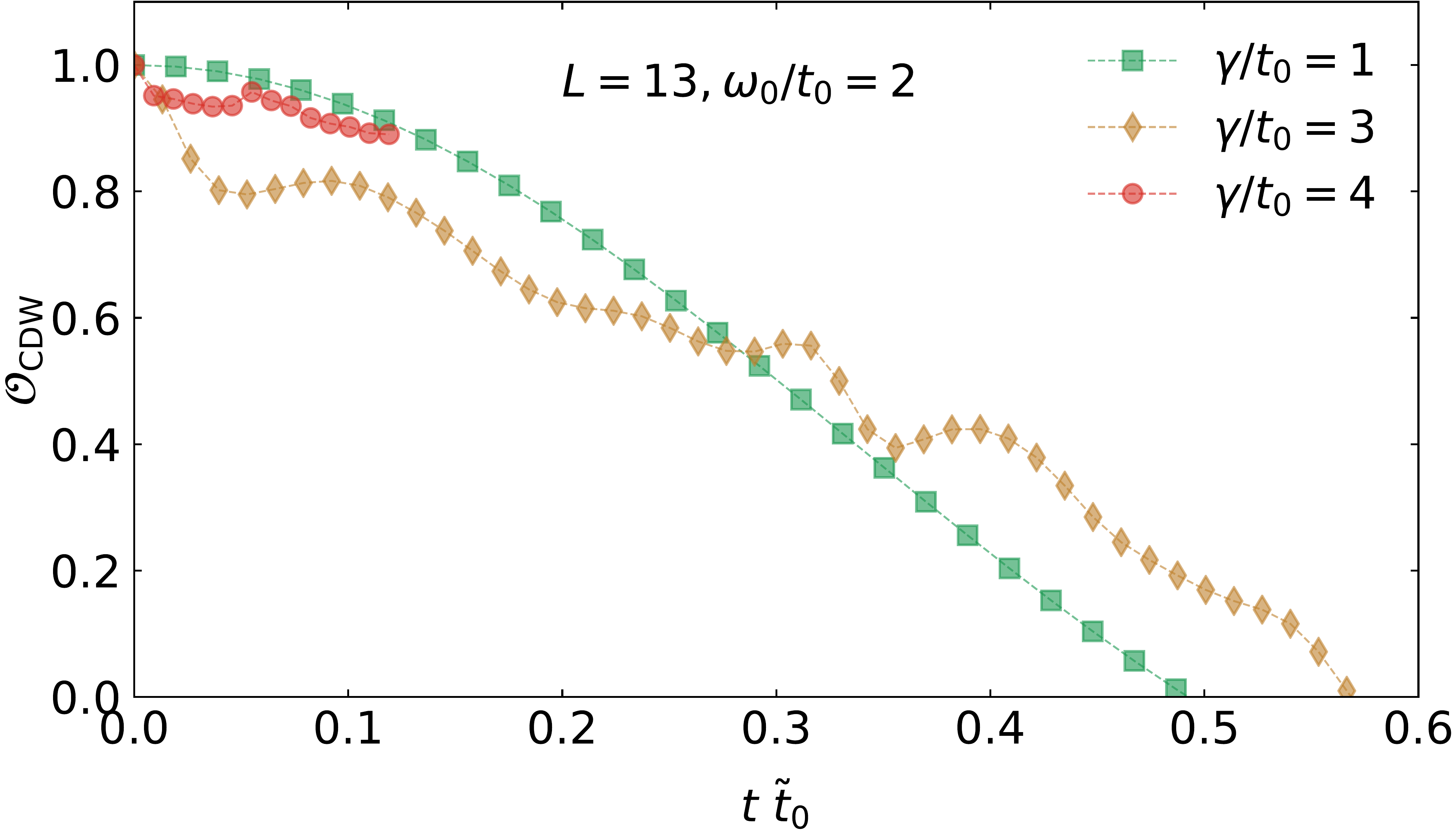}
\caption{\label{fig:dressed-Ocdw-t05-ttilde} Time evolution of the decay of the charge-density-wave order parameter $\mathcal{O}_{\rm CDW}$ when starting from the dressed CDW state $|\mathrm{DCDW}\rangle$ Eq.~\eqref{eq:DCDW}. Here, the time axis is in units of the effective hopping matrix element $\tilde t_0$ Eq.~\eqref{eq:teff} (system size $L=13$, phonon frequency $\omega_0/t_0 = 2$, and data for different coupling strengths $\gamma/t_0 = 1, 3, 4$). In the time evolution, we use a local phonon cutoff $M_{\rm ph} = 20, 30, 40$, respectively. The local discarded weight is set to $\Delta_{\rm loc} = 10^{-8}$. For clarity, we only show every fifth [every twentieth] data point that was computed for $\gamma/t_0 = 3$ [$\gamma/t_0 = 4$].}
\end{figure}

The different timescales of the dynamics in Fig.~\ref{fig:dressed-Ocdw-t05}(a) can also be understood in terms of decreasing effective hopping matrix elements for the polarons for increasing coupling strength $\gamma/t_0$. In Fig.~\ref{fig:dressed-Ocdw-t05-ttilde}, we plot the order parameter $\mathcal{O}_{\rm CDW}$ as a function of time where time is expressed in units of the inverse effective hopping matrix element $\tilde t_0$, Eq.~\eqref{eq:teff}, from the small-$t_0$ perturbation theory \cite{Hirsch-PRB-1983}. This does not produce a complete collapse of the data sets since we are already far away from the small-$t_0$ limit. Nevertheless, the decay of the order parameter now happens on comparable time scales for the different coupling strengths.

\begin{figure}[!tb]
\includegraphics[width=\columnwidth]{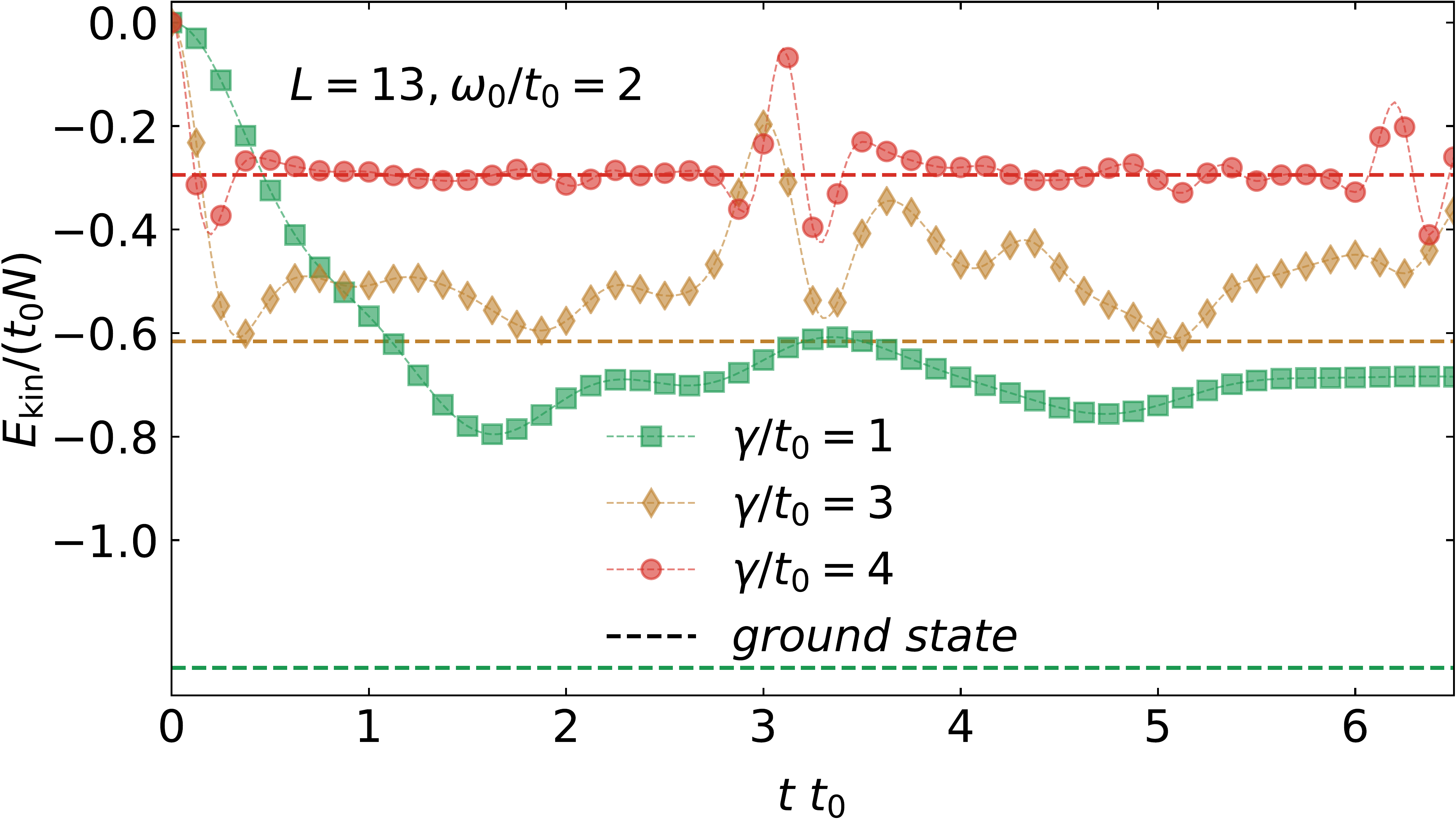}
\caption{\label{fig:dressed-Ekin-t05} Time evolution of the kinetic energy per fermion $E_{\rm kin}/N$ when starting from the dressed CDW state $|\mathrm{DCDW}\rangle$ Eq.~\eqref{eq:DCDW}. The dashed horizontal lines represent the kinetic energy in the ground states at the respective parameters. Simulations are performed for  $L=13$, $\omega_0/t_0 = 2$ and different coupling strengths $\gamma/t_0 = 1, 3, 4$. In the time evolution, we use a local phonon cutoff $M_{\rm ph} = 20, 30, 40$, respectively. The local discarded weight is set to $\Delta_{\rm loc} = 10^{-8}$. For clarity, we only show every fifth data point that was computed.}
\end{figure}

Another feature that is noticeable in Fig.~\ref{fig:dressed-Ocdw-t05}(a) is peaks in $\mathcal{O}_{\rm CDW}$ around $t t_0 \approx 3.1$ and $t t_0 \approx 6.3$ for $\gamma/t_0 = 4$.
The first peak is also visible for $\gamma/t_0 = 3$. The positions in time of these features coincide with multiples of the phonon period $2 \pi / \omega_0$.
This becomes evident when comparing data for different phonon frequencies $\omega_0/t_0$ (not shown here).
These features are also very prominent in Fig.~\ref{fig:dressed-Ekin-t05} where we plot the kinetic energy as a function of time when starting from the DCDW state.
For the strong coupling $\gamma/t_0 = 4$, the kinetic energy relaxes to the ground-state value (dashed red line in Fig.~\ref{fig:dressed-Ekin-t05}) after $tt_0 \approx 0.5$ and fluctuates around it.
Around $t t_0 \approx 3.1$, a peak appears that corresponds to the one seen in Fig.~\ref{fig:dressed-Ocdw-t05}(a).
After $tt_0 \approx 3.5$, the kinetic energy again fluctuates around the ground-state value before the second peak appears around  $t t_0 \approx 6.3$.
In contrast, the kinetic energy at $\gamma/t_0 = 1$ slowly decays to $E_{\rm kin}/(t_0 N) \approx -0.7$ and only shows very slow fluctuations around that value.
It is worth noting that this value is still far above the ground-state kinetic energy (horizontal dashed green line in Fig.~\ref{fig:dressed-Ekin-t05}).
The latter is not surprising since the initial DCDW state is far away from the ground state at these parameters. 

Comparing the time evolution of the BCDW state and the DCDW state, one notices that the behavior at the weak coupling $\gamma/t_0 = 1$ in the two cases is very similar.
The order parameter decays toward zero very fast and oscillates with a frequency controlled by the hopping parameter $t_0$. In contrast, the behavior for the stronger couplings $\gamma/t_0 = 3,4$ is quite different for the two different initial states.
When starting from the BCDW state the initial movement of the fermions is not affected much by the coupling to the phonons and only after a transient time, when phonons are emitted by the fermions and the polarons are formed, the fermions become very slow.
However, this slowing down of the movement is only temporary and after the phonons are reabsorbed the dynamics of the fermions speeds up again.
In contrast, the DCDW state at $\gamma/t_0 = 3,4$ already contains very heavy polarons and the movement of the fermions is slow right from the beginning.
A closely related behavior has been seen in a recent work by Kloss {\it et al.} \cite{Kloss-PRL-2019} in the expansion of a single particle injected into an empty Holstein lattice.
When the particle is initially dressed by phonons, the expansion is strongly suppressed as the coupling strength is increased.
In the opposite case of a bare electron, a repeated temporal suppression of the dynamics over time intervals of one phonon period is observed.
We find both these phenomena in the time evolution of the dressed and bare CDW state, respectively.

\begin{table}[b]
\caption{\label{tab:energies}Energy difference between the ground states and the initial states $\Delta E_{\mathrm{BCDW} [ \mathrm{DCDW} ]} = E_{\mathrm{BCDW}  [ \mathrm{DCDW} ]} - E^{\rm gs}$ for the BCDW [DCDW] state with $L = 13$ and $\omega_0/t_0 = 2$.}
\begin{ruledtabular}
\begin{tabular}{c c c}
$\gamma/t_0$ & $\Delta E_{\rm BCDW}/(t_0N)$ & $\Delta E_{\rm DCDW}/(t_0N)$ \\
\colrule
1  & 1.674 & 1.174 \\
3 & 4.877 & 0.377\\
4 & 8.153 & 0.153\\
\end{tabular}
\end{ruledtabular}
\end{table}

Another aspect is that the BCDW state as initial state is increasingly farther away from the ground state as $\gamma/\omega_0$ increases (cf. Table~\ref{tab:energies}).
On the other hand, in the case of the DCDW state the opposite is true.
The stronger the coupling $\gamma/\omega_0$ the closer the initial state is to the ground state in terms of energy.
This explains the slower relaxation due to the smaller fraction of intermediate states available in the many-body spectrum.

\subsection{\label{sec:quench}Quench from CDW to metallic phase}

In contrast to the initial CDW product states discussed in the previous sections, we now start from a fully correlated CDW state, i.e., the many-body ground state.
The quench protocol is as follows.
We prepare the system in the ground state for parameters in the CDW phase.
Then, at time $t = 0$, we quench the phonon frequency $\omega_0/t_0$ and the electron-phonon coupling parameter $\gamma/t_0$ such that for the resulting parameter set, the system is in the metallic TLL phase.
The quenches considered here are illustrated in the sketch of the phase diagram in Fig.~\ref{fig:holstein_sketch}(b) as arrows.
The horizontal arrow (FQ) illustrates the quench of both the phonon frequency $\omega_0/t_0$ and the coupling strength $\gamma/\omega_0$ in such a way that $\gamma/\omega_0$ stays constant, while the vertical arrow (CQ) illustrates the quench of only the coupling $\gamma/t_0$.
As mentioned earlier, we use an odd system size $L$ to pin the charge-density wave and get a nonzero value for the order parameter $\mathcal{O}_{\rm CDW}$ in the initial ground state. The number of particles in the system is then $N = (L-1)/2$.

In Table~\ref{tab:quench_energies}, we list the quench energies $\Delta E^{\rm qu} = E^{\rm init} - E^{\rm gs}$ in the two quenches, which is the difference between the energy of the state after the quench $E^{\rm init}$ and the ground-state energy $E^{\rm gs}$ for these parameters.
Furthermore, we list the kinetic and phonon quench energy $\Delta E_{\rm kin}^{\rm qu} = E_{\rm kin}^{\rm init} - E_{\rm kin}^{\rm gs}$ and $\Delta E_{\rm ph}^{\rm qu} = E_{\rm ph}^{\rm init} - E_{\rm ph}^{\rm gs}$ respectively, where $E_{\alpha} = \langle H_{\rm \alpha} \rangle$.
The kinetic part of the quench energy is very similar in the two quenches while the phononic part is not.
In the frequency quench, we reduce the energy of individual phonons with respect to the bandwidth and therefore, $\Delta E_{\rm ph}^{\rm qu}$ becomes quite small.
In comparison, in the coupling quench the ratio of phonon energy and bandwidth stays fixed and $\Delta E_{\rm ph}^{\rm qu}$ dominates $\Delta E^{\rm qu}$.
This explains why in the frequency quench, $\Delta E^{\rm qu}$ is smaller than in the coupling quench.

\begin{table}[b]
\caption{\label{tab:quench_energies} Total quench energies $\Delta E^{\rm qu}$ and the contributions from the kinetic part $\Delta E^{\rm qu}_{\rm kin}$ and the phononic part $\Delta E^{\rm qu}_{\rm ph}$.}
\begin{ruledtabular}
\begin{tabular}{l c c c c}
\ & $\Delta E^{\rm qu}/(t_0N)$ & $\Delta E^{\rm qu}_{\rm kin}/(t_0N)$  & $\Delta E^{\rm qu}_{\rm ph}/(t_0N)$ \\
\colrule
FQ & 0.822 & 1.007 & 0.197 \\
CQ & 5.190 & 0.952 & 7.398 \\
\end{tabular}
\end{ruledtabular}
\end{table}

\subsubsection{\label{sec:FQ} Frequency quench}

\begin{figure}[!tb]
\includegraphics[width=\columnwidth]{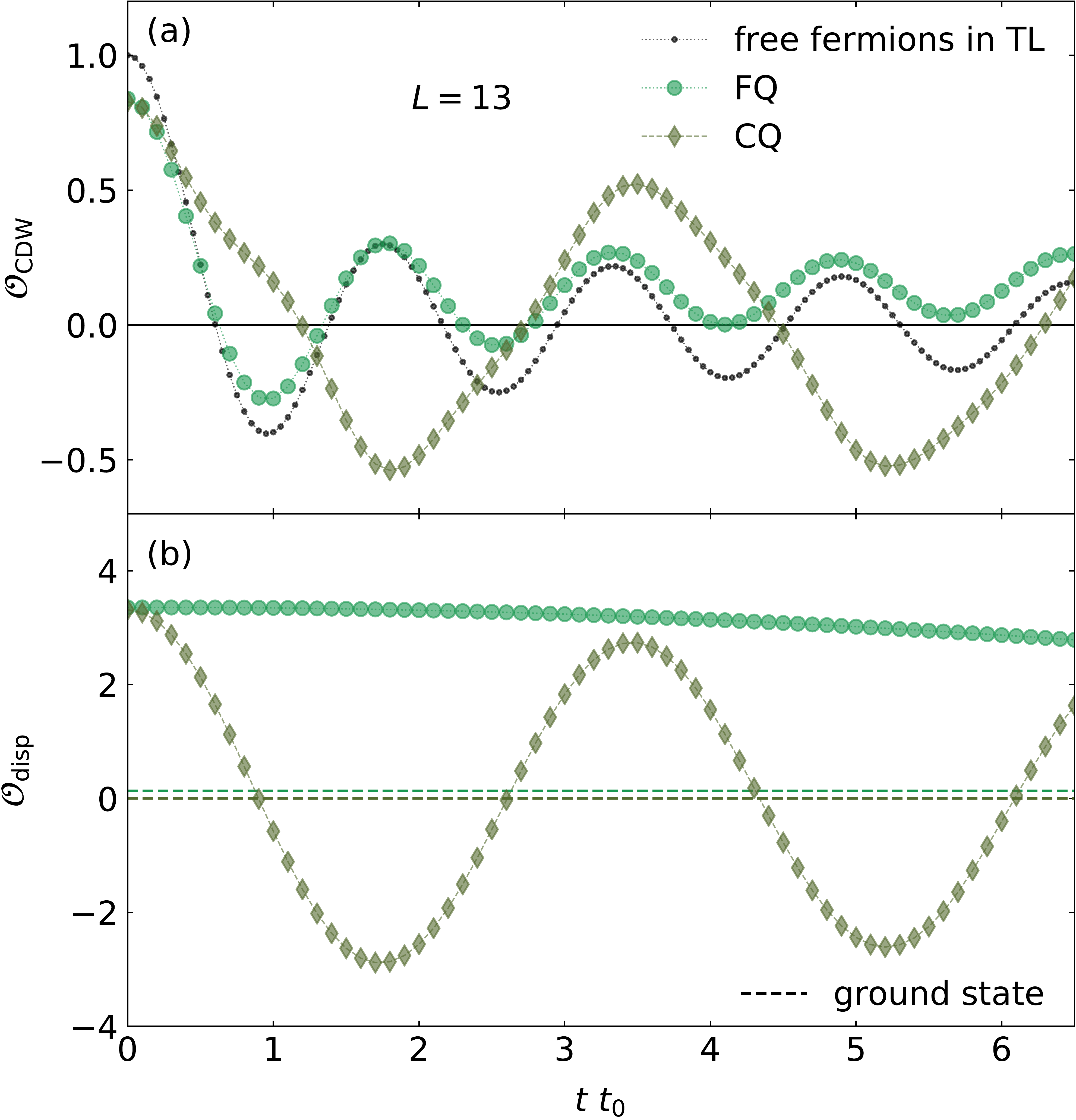}
\caption{\label{fig:g_quench_Ocdw} Time evolution of (a) the decay of the charge-density-wave order parameter $\mathcal{O}_{\rm CDW}$ and (b) the staggered displacement $\mathcal{O}_{\rm disp}$ Eq.~\eqref{eq:Ocdwph} in a quench from the CDW phase to the TLL phase. Circles: Quench from $\omega_{0, \rm init}/t_0 = 2$ and $\gamma_{\rm init}/t_0 = 4$ to $\omega_0/t_0 = 0.1$ and $\gamma/t_0 = 0.2$ (frequency quench, FQ). Diamonds: Quench from $\gamma_{\rm init}/t_0 = 4$ to $\gamma/t_0 = 1$ while $\omega_0/t_0 = 2$ is kept fixed (coupling quench, CQ). The small black dots in panel (a) are exact analytical results in the thermodynamic limit for $| \mathrm{BCDW} \rangle$ as initial state and no coupling to phonons \cite{Barmettler-PRL-2009}. The dashed horizontal lines in panel (b) represent the value of $\mathcal{O}_{\rm disp}$ in the respective ground states. The system size is $L = 13$, local phonon cutoff $M_{\rm ph} = 40$, and the local discarded weight is set to $\Delta_{\rm loc} = 10^{-8}$. For clarity, we only plot every second [fourth] data point that was computed in the FQ [CQ].}
\end{figure}

We first consider the frequency quench, starting from the ground state at  $\omega_{0, \rm init}/t_0 = 2$ and $\gamma_{\rm init}/t_0 = 4$ which is in the CDW phase.
At $t=0$, we quench the phonon frequency to $\omega_0/t_0 = 0.1$ and the coupling strength to $\gamma/t_0 = 0.2$  with $\gamma/\omega_0=2=$const.
The time evolution of the order parameter is shown in Fig.~\ref{fig:g_quench_Ocdw}(a) as circles.
The order quickly decays toward zero and oscillates around a value slightly bigger than zero.
The comparison to the exact analytical results for relaxation from the BCDW state with $\gamma = 0$ \cite{Barmettler-PRL-2009} [small black dots in Fig.~\ref{fig:g_quench_Ocdw}(a)] reveals that the frequency of the oscillations is controlled by the hopping parameter $t_0$.
Moreover, the electrons clearly move into the previously empty sites.

Instead of the phonon number, we discuss the staggered displacement to characterize the dynamics in the 
phonon sector:
\begin{align}
\mathcal{O}_{\rm disp} = \frac{1}{N} \sum_{l = 1}^L (-1)^l  \langle b_l^\dagger + b_l^{\vphantom{\dagger}} \rangle \,. \label{eq:Ocdwph}
\end{align}
$\langle b_l^\dagger + b_l^{\vphantom{\dagger}} \rangle$ is the expectation value of the displacement of the harmonic oscillator on site $l$. In equilibrium, a nonzero value of the fermion CDW order parameter $\mathcal{O}_{\rm CDW}$ is accompanied by a nonzero value of the staggered displacement $\mathcal{O}_{\rm disp}$. 
We plot the staggered displacement in  Fig.~\ref{fig:g_quench_Ocdw}(b).
For the FQ, it remains positive during the simulation window and decreases only slightly. 
To qualitatively understand the nonequilibrium phenomena investigated here it is helpful to adapt a mean-field-like picture.
The displacements of the harmonic oscillators can be viewed as a potential landscape for the electrons when we replace the displacement operators in Eq.~\eqref{eq:ham-el-ph} by their expectation values.

\begin{figure}[!tb]
\includegraphics[width=\columnwidth]{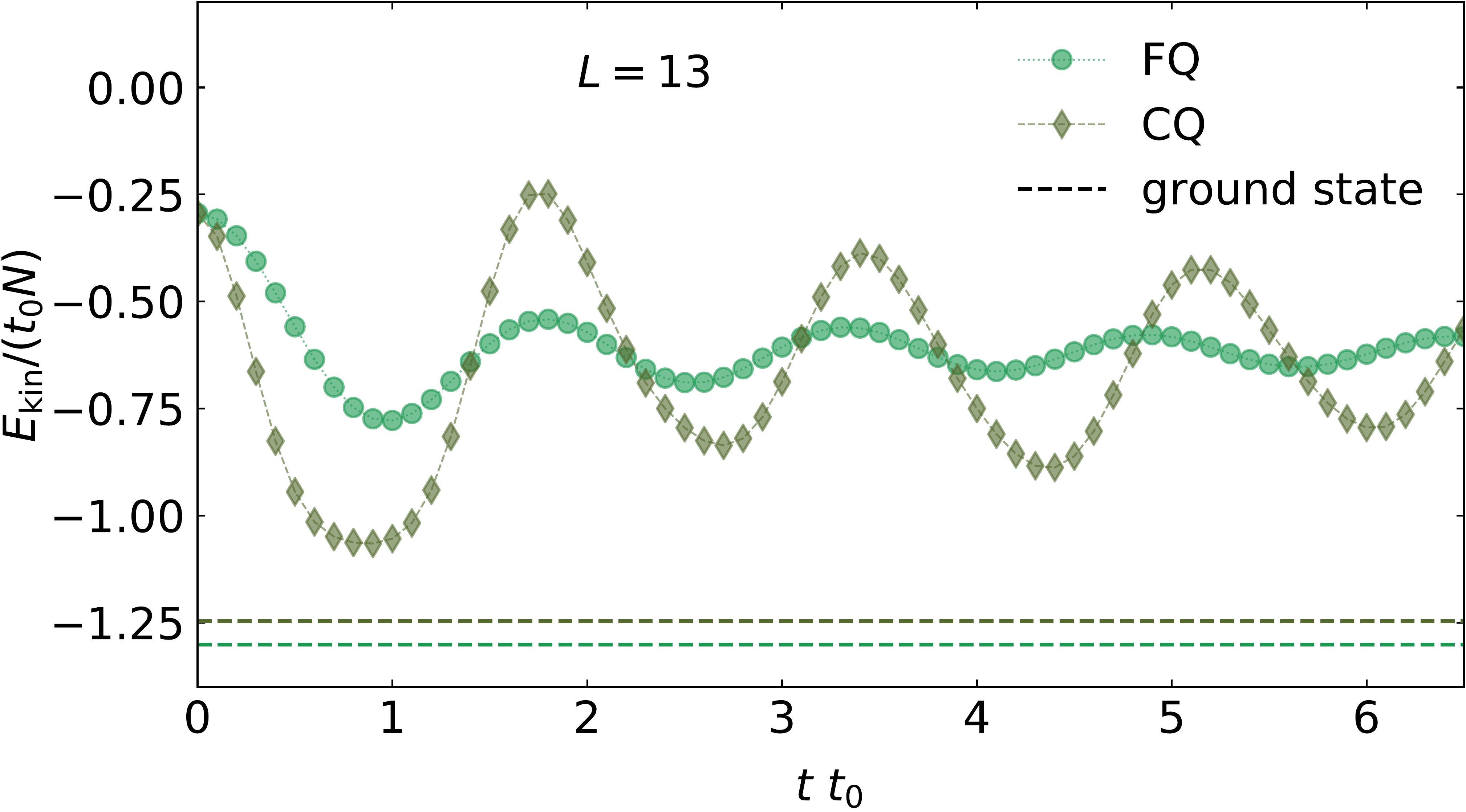}
\caption{\label{fig:g_quench_Ekin} Time evolution of the kinetic energy per fermion $E_{\rm kin}/N$ in a quench from the CDW phase to the TLL phase. Circles: Quench from $\omega_{0, \rm init}/t_0 = 2$ and $\gamma_{\rm init}/t_0 = 4$ to $\omega_0/t_0 = 0.1$ and $\gamma/t_0 = 0.2$ (frequency quench, FQ). Diamonds: Quench from $\gamma_{\rm init}/t_0 = 4$ to $\gamma/t_0 = 1$ while $\omega_0/t_0 = 2$ is kept fixed (coupling quench, CQ). The dashed horizontal lines represent the kinetic energy per fermion $E_{\rm kin}/N$ in the respective ground states. The system size is $L = 13$, local phonon cutoff $M_{\rm ph} = 40$, and the local discarded weight is set to $\Delta_{\rm loc} = 10^{-8}$. For clarity, we only plot every second [fourth] data point that was computed in the FQ [CQ].}
\end{figure}

In the case of the frequency quench, the staggered displacement $\mathcal{O}_{\rm disp}$ changes very slowly as a function of time since the phonon period $2\pi/\omega_0$ is very large.
As a consequence, at the end of our simulation time, there is still a background potential landscape.
The electrons move in  this background potential and therefore, their order remains larger than in the free case $\gamma=0$. 
This also means that although the electron CDW order parameter $\mathcal{O}_{\rm CDW}$ exhibits a fast dynamics and only shows small oscillations, the entire system is still very far from equilibration since the phonons remain in a spatially inhomogeneous state. 
In  order to observe the relaxation of the whole system towards a stationary state one would have to simulate to much longer times than what is currently feasible with our method. Finally, in Fig.~\ref{fig:g_quench_Ekin} we present the kinetic energy after the frequency quench. It relaxes towards an almost stationary value after $t t_0 \approx 1.5$ with only small oscillations with a frequency similar to that in the time evolution of $\mathcal{O}_{\rm CDW}$.

\subsubsection{\label{sec:CQ} Coupling quench}

In the second quench scenario, we fix the phonon frequency to $\omega_0/t_0 = 2$ and quench only the coupling strength from $\gamma_{\rm init}/t_0 = 4$ to $\gamma/t_0 = 1$. The time evolution of the order parameter $\mathcal{O}_{\rm CDW}$ is plotted as diamonds in Fig.~\ref{fig:g_quench_Ocdw}(a).
In contrast to the frequency quench, the order parameter in the coupling quench shows large slow oscillations with an amplitude that barely decreases on the timescales that are accessible here.
In Fig.~\ref{fig:g_quench_Ocdw}(b), we plot the staggered displacement $\mathcal{O}_{\rm disp}$ in this quench as diamonds. One can see that the staggered displacement oscillates with a period of $2\pi/\omega_0$ between positive and negative values; i.e., the phonons, once released from the polaron, start to undergo a nonequilibrium dynamics with oscillating displacement.
Note that the phonon density itself also remains largely concentrated on the even sites (data not shown here).
For the effective potential landscape this means that the fermions are attracted to their initial places when $\mathcal{O}_{\rm disp}$ is positive and are pushed away from these sites when $\mathcal{O}_{\rm disp}$ is negative.
Therefore, the oscillations in  $\mathcal{O}_{\rm CDW}$ and $\mathcal{O}_{\rm disp}$ are locked to one another and the frequencies are comparable.
Similar to the FQ, the spatially inhomogeneous nonequilibrium distribution of the phonons remains stable.

This locking effect also explains the oscillations in the kinetic energy plotted as diamonds in Fig.~\ref{fig:g_quench_Ekin}.
The kinetic energy has a maximum whenever $\mathcal{O}_{\rm disp}$ has a maximum or a minimum. 
This occurs when the fermions are localized on the even or odd sites, respectively.
On the other hand, when the potential landscape is closer to being flat and $\mathcal{O}_{\rm disp}$ is close to zero, the fermions hop around and the kinetic energy has a minimum.

In summary, the quenches again exhibit strong dependencies on the initial state and on the final-state parameters in the transient dynamics.
As in the relaxation dynamics of the BCDW and DCDW states, the phonons primarily slow down the electronic dynamics.
For the postquench parameters in the TLL phase considered here, the electrons can move but the phonon distribution relaxes much slower, resulting in a slowly decaying
inhomogeneous nonequilibrium distribution. 
It would be very interesting to extend the analysis to the case of dispersive phonons to study whether this can speed up both the electronic relaxation and the dissolving of spatially inhomogeneous phonon distributions.
From a broader perspective, this leads to the topic of energy transport, which in the Holstein model can only occur via electronic quasiparticle motion while dispersive phonons could carry an energy current themselves.
These questions are left for future studies.

\section{Summary}\label{sec:summary}

To summarize, we studied the melting of CDW order by means of real-time simulations of the half-filled Holstein model of spinless fermions in one dimension.
To this end, we investigated relaxation dynamics that is dominated by electron-phonon coupling in the far-from-equilibrium regime, complementary to the case studied in \cite{Hashimoto-PRB-2017} where strong electron interactions were present.
We find a strong dependence of the transient dynamics on the precise initial state and on the model parameters.
As discussed in previous work \cite{Golez-PRL-2012,Matsueda-JPSJ-2012,Hashimoto-PRB-2017,Mendoza-Arenas-PRB-2019,Kloss-PRL-2019}, a main effect of an electron-phonon coupling is the slowing down of the dynamics of the electrons compared to a purely electronic system.
This is attributed to the formation of polarons which renormalizes the mass of the charge carriers.
For weak coupling the movement of the electrons is comparable to the dynamics of free particles with small corrections.
In the case of strong coupling, the dynamics on transient timescales can be altered more drastically, which is exemplified by the temporal self-trapping of the electrons observed here.

Furthermore, we often find very different timescales for the relaxation in the electron and the phonon sector as is most clearly evident in the quenches from correlated ground states.
In these situations, we observe that the initial spatially inhomogeneous phonon distribution persists and forms a potential background for the electron relaxation.
As a result, inhomogeneities remain in the spatial electron distribution as well.
A question for further studies is how this picture changes when introducing a dispersion of the phonons.
It remains as an open question whether regimes can be found where the presence of phonons actually accelerates the full relaxation of the electronic system.
This connects our work to the question of how inhomogeneities in the phonon sector of an electron-phonon coupled system relax and, more generally, how different channels of energy and charge transport compete in such systems (in the context of the Su-Schrieffer-Heeger model, such questions were discussed in, e.g., \cite{Mendoza-Arenas-PRB-2019}).

Our work demonstrates the capabilities of combining LBO with MPS-based numerical methods when applied to electron-phonon coupled systems.
The TEBD-LBO algorithm gives access to regimes far from equilibrium that are out of reach for conventional MPS-based techniques.
We postpone the question of a benchmark of our DMRG3S+LBO algorithm against other state-of-the-art ground-state DMRG algorithms that were developed for electron-phonon coupled systems (such as the pseudosite method \cite{Jeckelmann-PRB-1998}) to future studies.

\begin{acknowledgments}
We acknowledge useful discussions with J. Bon\v ca, C. Brockt, C. Hubig, E. Jeckelmann, and L. Vidmar. This work was supported by the Deutsche Forschungsgemeinschaft (DFG, German Research Foundation) under Project No. 207383564 via Research Unit FOR 1807 and via SFB 1073 (project B09) under Project No. 217133147. J.H. acknowledges support from the Polish National Agency of Academic Exchange (NAWA) under Contract No. PPN/PPO/2018/1/00035. E.D. was supported by the US Department of Energy (DOE), Office of Science, Basic Energy Sciences (BES), Materials Sciences and Engineering Division.
\end{acknowledgments}

\appendix

\section{Finite-size dependence}

\begin{figure}[!hb]
\includegraphics[width=\columnwidth]{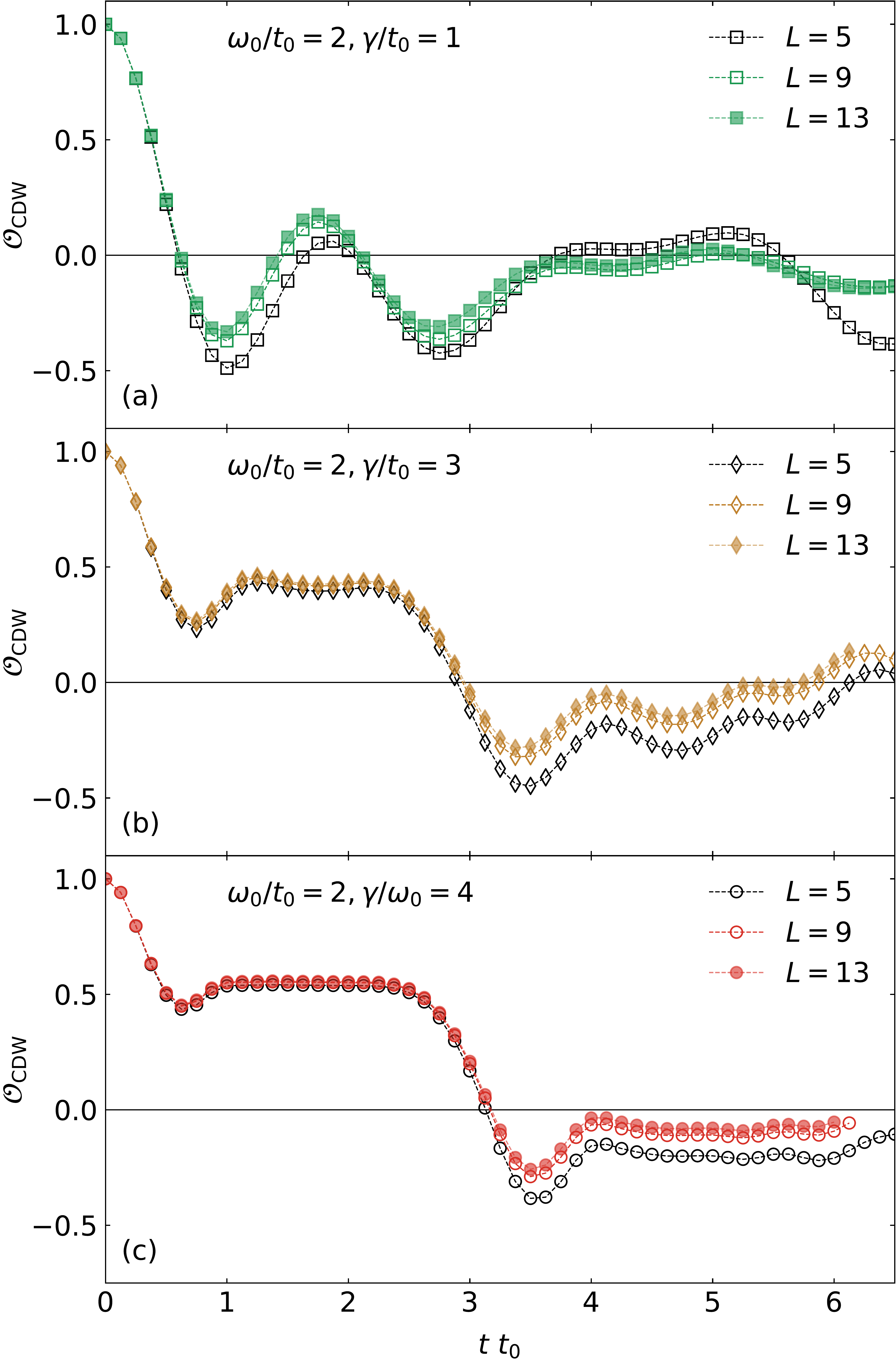}
\caption{\label{fig:bare_Ocdw_sys} Time evolution of the charge-density-wave order parameter $\mathcal{O}_{\rm CDW}$ when starting from the bare CDW state $|\mathrm{BCDW}\rangle$ Eq.~\eqref{eq:BCDW}. Comparison between different system sizes $L = 5$ (open black symbols), $L = 9$ (open colored symbols) and $L = 13$ (filled symbols). The phonon frequency is $\omega_0/t_0 = 2$ while in panel (a) $\gamma/t_0 = 1$, in panel (b) $\gamma/t_0 = 3$ and in panel (c) $\gamma/t_0 = 4$. In the time evolution, we use a local phonon cutoff $M_{\rm ph} = 10, 30, 40$, respectively, and the local discarded weight is set to $\Delta_{\rm loc} = 10^{-8}$. For clarity, we only show every fifth data point that was computed.}
\end{figure}

\begin{figure}[!tb]
\includegraphics[width=\columnwidth]{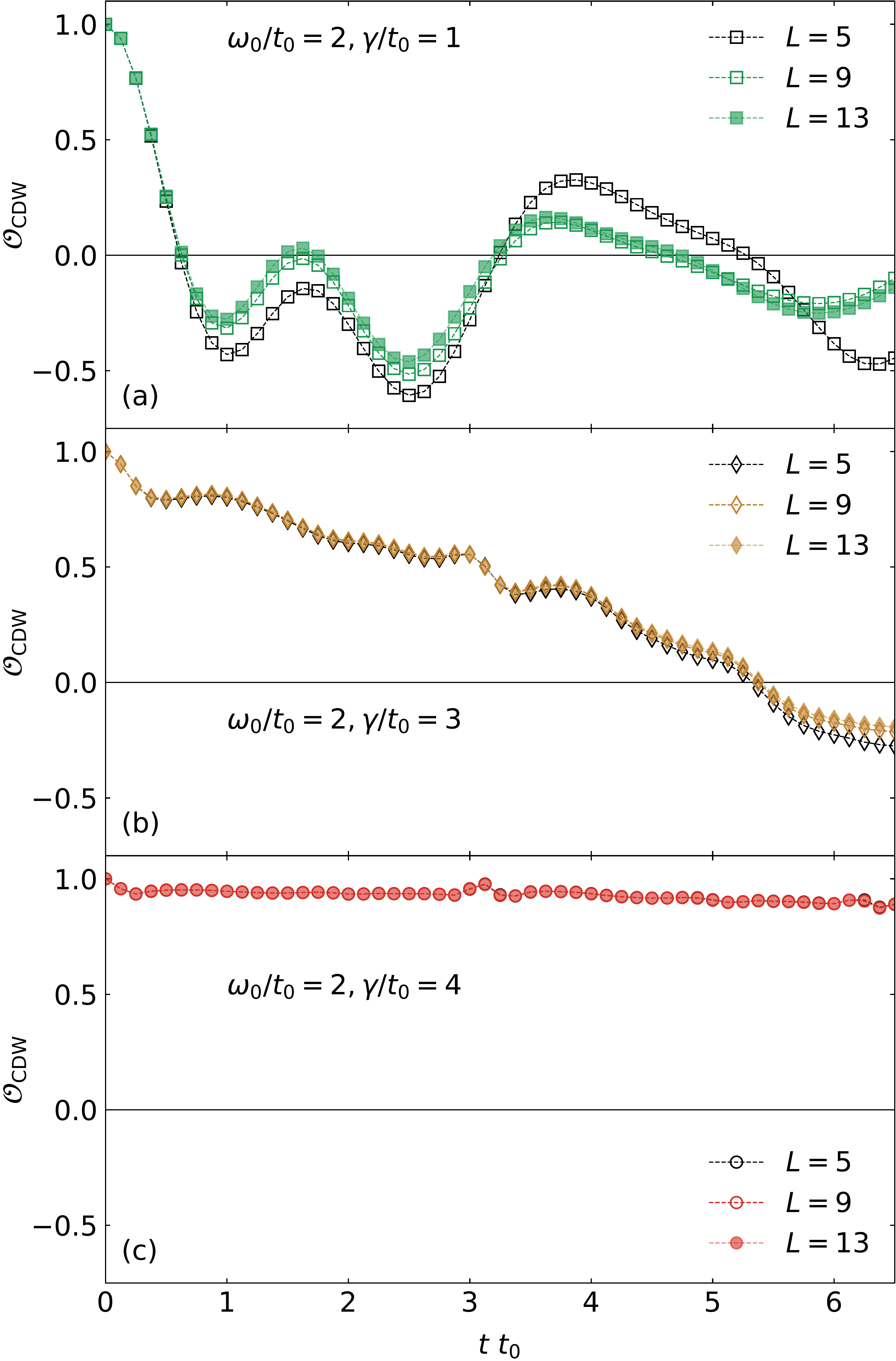}
\caption{\label{fig:dressed_Ocdw_sys} Time evolution of the charge-density-wave order parameter $\mathcal{O}_{\rm CDW}$ when starting from the dressed CDW state $|\mathrm{DCDW}\rangle$ Eq.~\eqref{eq:DCDW}. Comparison between different system sizes $L = 5$ (open black symbols), $L = 9$ (open colored symbols) and $L = 13$ (filled symbols). The phonon frequency is $\omega_0/t_0 = 2$ while in panel (a) $\gamma/t_0 = 1$, in panel (b) $\gamma/t_0 = 3$ and in panel (c) $\gamma/t_0 = 4$. In the time evolution, we use a local phonon cutoff $M_{\rm ph} = 20, 30, 40$, respectively, and the local discarded weight is set to $\Delta_{\rm loc} = 10^{-8}$. For clarity, we only show every fifth data point that was computed.}
\end{figure}

\begin{figure}[!tb]
\includegraphics[width=\columnwidth]{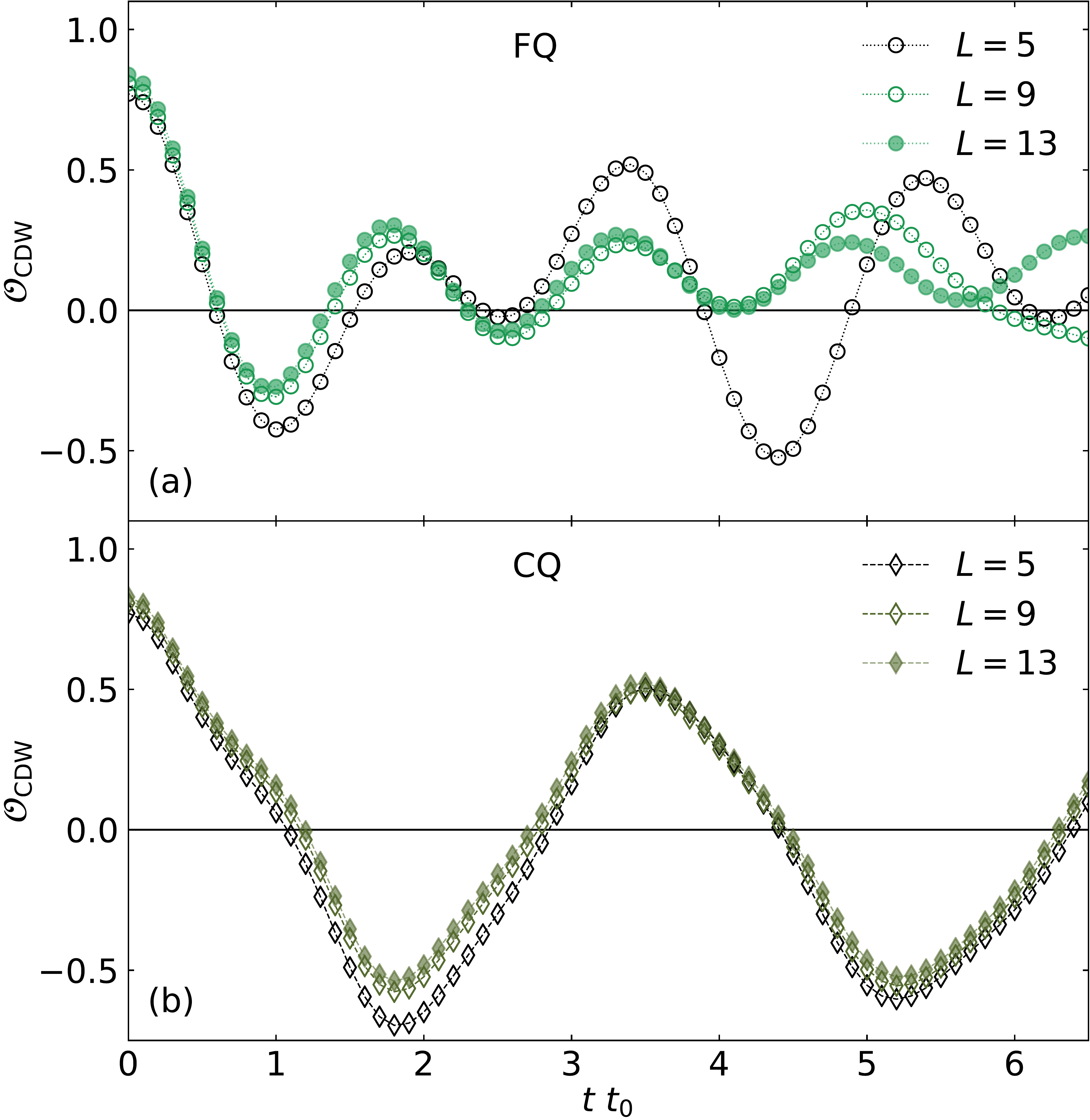}
\caption{\label{fig:quench_Ocdw_sys} Time evolution of the charge-density-wave order parameter $\mathcal{O}_{\rm CDW}$ for different system sizes $L = 5$ (open black symbols), $L = 9$ (open colored symbols) and $L = 13$ (filled symbols). Panel (a): quench from $\omega_{0, \rm init}/t_0 = 2$ and $\gamma_{\rm init}/t_0 = 4$ to $\omega_0/t_0 = 0.1$ and $\gamma/t_0 = 0.2$ (frequency quench). Panel (b): quench from $\gamma_{\rm init}/t_0 = 4$ to $\gamma/t_0 = 1$ while $\omega_0/t_0 = 2$ is kept fixed (coupling quench). The local phonon cutoff is $M_{\rm ph} = 40$ and the local discarded weight is set to $\Delta_{\rm loc} = 10^{-8}$. For clarity, we only plot every second data point that was computed in panel (a) and every fourth data point that was computed in panel (b).}
\end{figure}

In Figs.~\ref{fig:bare_Ocdw_sys}, \ref{fig:dressed_Ocdw_sys}, and \ref{fig:quench_Ocdw_sys}, we compare time-evolution data produced with our TEBD-LBO method (cf. Sec.~\ref{sec:tebd+lbo}) for different system sizes $L = 5,9,13$.

In Fig.~\ref{fig:bare_Ocdw_sys}, the initial state is the bare CDW state and the phonon frequency is set to $\omega_0/t_0 = 2$ as in Sec.~\ref{sec:bare}.
The largest finite-size effects are seen for $\gamma/t_0 = 1$ in Fig.~\ref{fig:bare_Ocdw_sys}(a).
This is expected since because of the weak coupling, the dynamics is the fastest here.
Nevertheless, there are no big qualitative differences between the different system sizes.
Finite-size effects are even smaller for the larger couplings $\gamma/t_0 = 3$ [Fig.~\ref{fig:bare_Ocdw_sys}(b)] and $\gamma/t_0 = 4$ [Fig.~\ref{fig:bare_Ocdw_sys}(c)] where until $tt_0 \approx 2.8$, the data of the different system sizes lie on top of each other and only small deviations are seen for larger times.

The picture is similar for the DCDW state (cf. Sec.~\ref{sec:dressed}) as the initial state. In Fig.~\ref{fig:dressed_Ocdw_sys}, we compare the time evolution of $\mathcal{O}_{\rm CDW}$ at $\omega_0/t_0 = 2$ for the system sizes $L = 5,9,13$. Again, the largest finite-size effects are seen in panel (a) of Fig.~\ref{fig:dressed_Ocdw_sys} for $\gamma/t_0 = 1$. For $\gamma/t_0 = 3$ [Fig.~\ref{fig:dressed_Ocdw_sys}(b)] small finite-size effects are observable for $tt_0 \gtrsim 4$, while for $\gamma/t_0 = 4$ [Fig.~\ref{fig:dressed_Ocdw_sys}(c)] the data for the different system sizes lie on top of each other for the full simulation time. This is a manifestation of the very slow dynamics and the proximity to the ground state of the dressed CDW state at large $\gamma/t_0$.

In Fig.~\ref{fig:quench_Ocdw_sys}, we compare different system size data for the quenches discussed in Sec.~\ref{sec:quench}.
In Fig.~\ref{fig:quench_Ocdw_sys}(a), the phonon frequency and coupling strength are quenched from $\omega_{0,\rm init}/t_0 = 2$ and $\gamma_{\rm init}/t_0 = 4$ to $\omega_{0}/t_0 = 0.1$ and $\gamma/t_0 = 0.2$.
Here, we can observe large boundary effects for $L = 5$ after $tt_0 \approx 3$ and for $L = 9$ after $tt_0 \approx 4.7$.
This is not surprising since the dynamics is dominated by the hopping parameter $t_0$ in this case as discussed in Sec.~\ref{sec:quench}.
The largest velocity in the system is therefore $v_{\rm max} \approx 2t_0$ and hence the fastest excitations had time to travel across the entire system and bounce back from the boundary.

The situation is different in Fig.~\ref{fig:quench_Ocdw_sys}(b) for the quench of the coupling strength $\gamma_{\rm init}/t_0 = 4$ to $\gamma/t_0 = 1$ while the phonon frequency is fixed to $\omega_0/t_0 = 2$.
In this case, the finite-size effects seen are very small which is evidence for the fact that the dynamics in the system is not dominated by the free movement of the electrons.
Instead, the presence of the phonons from the CDW initial state plays the key role in the dynamics.

Overall, Figs.~\ref{fig:bare_Ocdw_sys}, \ref{fig:dressed_Ocdw_sys}, and \ref{fig:quench_Ocdw_sys} show that the key features in the time evolution of $\mathcal{O}_{\rm CDW}$ that are described in Sec.~\ref{sec:nonequ} are robust against finite-size effects and are not just an effect of the small system sizes considered in this work.

\clearpage
\bibliography{biblio-holstein-cdw}

\end{document}